\documentclass{aa}  
\usepackage[varg]{txfonts}
\usepackage{subcaption}

\usepackage{threeparttable}
\usepackage{url}

\usepackage[breaklinks=true]{hyperref}

\title{Analysis of a coronal hole jet in the chromosphere and the transition region with IRIS and SDO observations}
\titlerunning{Analysis of a CH jet in the chromosphere and TR with IRIS and SDO}

\author{
  M. Koletti\inst{1,2}
  \and
  C. Gontikakis\inst{2}
  \and
  S. Patsourakos\inst{3}
  \and
  K. Tsinganos\inst{1}
}

\institute{
  Department of Physics, National and Kapodistrian University of Athens, University Campus, Zografos, GR-157 84 Athens, Greece
  \and
  Research Center for Astronomy and Applied Mathematics, Academy of Athens, Soranou Efesiou Str., 4, 11527, Athens, Greece
  \and
  Physics Department, University of Ioannina, Ioannina 45110, Greece
}

\begin{document}

\abstract
{
Coronal jets are transient phenomena that occur throughout the solar atmosphere, and they are often triggered by the activation of mini-filaments. Jets are an important subject to study, as they are believed to contribute to the mass supply of the solar wind.
}
{
We studied a small-scale solar jet occurring in a low-latitude coronal hole, utilizing chromospheric and transition region diagnostics to investigate traces of the jet eruption in the lower atmosphere. We also investigate the possible mechanisms and dynamic conditions leading to its onset.
}
{
We analyzed ultraviolet spectra from the Interface Region Imaging Spectrograph (IRIS) to investigate chromospheric dynamics and their connection to the transition region, using diagnostics of the Mg~\textsc{ii} k 2796.35~\AA\ and Si~\textsc{iv} 1393.75~\AA\ lines together with non- local thermodynamic equilibrium inversions. We complemented this analysis with multiwavelength imaging and magnetograms from the Solar Dynamics Observatory (SDO) as well as magnetic field modeling.
}
{Our analysis of the Mg~\textsc{ii} k and Si~\textsc{iv} parameters revealed a coupling between chromospheric and transition region plasma dynamics. Profiles on the jet spire present deeper central reversals, larger peak separations, and higher transition region Doppler velocities. 
The jet initiation is associated with a mini-filament-like structure, which is preceded by dynamic variations in the photospheric magnetic field. We find single-peaked Mg \textsc{ii} k profiles at one of its footpoints, possibly indicating plasma drainage. The presence of a null point and a quasi-separatrix layer suggests that reconnection can play a role in the mini-filament-like structure destabilization and jet eruption. 
}
{
This study suggests that the signature and possibly the origin of coronal jets can be traced down to the chromosphere. The detailed analysis of the Mg \textsc{ii} k profiles at the jet spire and the mini-filament-like structure may provide valuable observational constraints for future magnetohydrodynamic (MHD) simulations. 
}


\maketitle

\section{Introduction}

Solar jets are transient, impulsive energy-release events that occur frequently throughout the solar atmosphere. Understanding the physical mechanisms that drive their formation is essential, as they are believed to contribute both to the heating of the upper atmosphere and mass supply of the solar wind \citep{Brueckner_1983,Chitta_2025}.
Observations of solar jets are commonly performed in the ultraviolet (UV) \citep{Shen_2021}, using high-resolution imaging from the Atmospheric Imaging Assembly \citep[AIA;][]{Lemen2012} and line-of-sight (LOS) magnetograms from the Helioseismic Magnetic Imager \citep[HMI;][]{Scherrer_2012}, both on board the Solar Dynamics Observatory \citep[SDO;][]{Pesnell_2012}, together with spectroscopic measurements from instruments such as the Interface Region Imaging Spectrograph \citep[IRIS;][]{DePontieu_2014}. 

A comprehensive understanding of these events requires combined observations that probe multiple atmospheric layers. IRIS provides several spectral lines suited for this purpose. The Si~\textsc{iv} 1393.755~\AA\ and 1402.770~\AA\ doublet lines are key emission lines for studying the dynamics of the solar transition region \citep{Innes_1997,Gontikakis_2018}. Moreover, the optically thick  Mg~\textsc{ii} k (2796.35~\AA) and h (2803.53~\AA) resonance lines (wavelengths taken from CHIANTI \citep{Del_Zanna_2021}) sample plasma conditions from the photosphere to the upper chromosphere and thus provide powerful diagnostics on the lower-atmospheric dynamics \citep{Lemaire_1981, Leenaarts_2013a, Leenaarts_2013b}. 

In quiet-Sun regions, Mg~\textsc{ii} h\&k profiles are typically double-peaked with a central reversal. However, there are many cases where they are single-peaked, for example in sunspot umbrae \cite{Kneer_1981,Bose_2019}, where the source function is never well coupled to the local temperature, due to the low density \citep{Hansteen_2023}. Single-peaked profiles in plages can result from a hot and dense chromosphere or a transition region moved to a higher column mass \citep{Carlsson_2015}. In flares, large asymmetries and the absence of a central reversal are presumably caused by increased temperatures and electron densities \citep{Lemaire_1984,Kerr_2015,Zhu_2019} or significant unresolved upflows and downflows \citep{Costa_Kleint_2017}. Fainter single-peaked profiles have also been found in filaments and specifically in low temperature regions \citep{Santos_2022}. However, interpreting these shapes in prominences has been proven nontrivial \citep{Heinzel_2014}. \cite{Barczynski_2023} found single-peaked profiles at the bases of coronal cavities in prominences, where plasma condensation or drainage toward the chromosphere is believed to take place.  Detailed observations of Mg \textsc{ii} h$\&$k profiles in mini-filaments are less common \citep{Madjarska_2022,Panesar_2022}. 
Asymmetric profiles are generally the result of velocity gradients, which are related to plasma motions above the peak formation height \citep{Leenaarts_2013b} and wave propagation \citep{Gouttebroze_1989}. The effects of microturbulence and opacity can also influence the width and shape of the profiles \citep{Athay_Skumanich_1968,Milkey_1974,Carlsson_2015}. However, the extent of turbulence of any scale in the chromosphere is still under study \citep{Judge_2020}.

Eruptions of mini-filaments are generally believed to be one of the primary progenitors of solar jet eruptions  \citep{Sterling_2015}. Such structures are miniature versions of large-scale filaments and typically lie above polarity inversion lines \citep{Wang_2000}.  
Observations frequently reveal that components of the jet spire appear at different wavelengths and temperatures, and are not strictly co-spatial \citep{Canfield_1996,Alexander_1999}. These components correspond to cool and hot plasma flows guided along different magnetic field lines but linked to the same eruptive event \citep{Chae_1999}. 

Magnetic reconnection is believed to be the fundamental driver of solar jets \citep{Shibata_1992,Archontis_2010,Pontin2022}. Spectroscopic evidence of reconnection can be bi-directional flows, such as downflows at the Mg~\textsc{ii} k line and upflows at the Si~\textsc{iv} line \citep{Gorman2022}. 
Magnetic reconnection can occur at null points, where the magnetic field vanishes, or within quasi-separatrix layers (QSLs), where magnetic connectivity changes rapidly but remains continuous \citep{Priest_Demoulin_1995,Titov_2002}. Studies have shown that jets often originate in null-point fan–spine topologies \citep{Moreno_Insertis_2008, Torok_2009,Kumar_2019}. Moreover, jets are also associated with reconnection in QSLs \citep{Li_2025}, even in the absence of null points \citep{Guo_2013}. 

Flux emergence and flux cancellation are commonly connected to solar eruptions \citep{Wyper_2017,Kliem_2005}. Flux emergence has been associated with filament formation and jet activity \citep{Chen_2018,Zhuleku_2025,Archontis_2019,Chen_2025_mf}, while flux cancellation is frequently observed prior to mini-filament eruptions and coronal jets \citep{Hong_2011,Panesar_2016}, leading to the disappearance of both or one of the polarities at the polarity inversion line \citep{Livi_1985, Martin_1988}.

This study significantly extends the work presented in \citeauthor{Koletti_2024} (\citeyear{Koletti_2024}; hereafter Paper I). Paper I focused on the transition region and coronal signatures of two small-scale jets, using Si~\textsc{iv} and AIA diagnostics. Here, we provide a detailed spectroscopic analysis of one of the jets (jet 1) in the chromosphere using the optically thick Mg~\textsc{ii} k line. Moreover, by combining the Si~\textsc{iv} diagnostics from Paper I with new constraints from Mg~\textsc{ii} k, we bridge chromospheric and transition region layers. This multi-height approach is essential for a comprehensive understanding of jet initiation and evolution, and it highlights the central role of mini-filament dynamics in driving solar jets. 
Multi-height spectroscopic studies of jets and mini-filaments, particularly those including chromospheric diagnostics, remain scarce (e.g., \citealt{Joshi_2020, Cai_2019, Panesar_2025, Lu_2019, Gorman2022, Madjarska_2022}). In this context, our work represents one of the few studies that combines high-resolution chromospheric and transition region spectroscopy in order to probe jet dynamics across multiple atmospheric layers while specifically focusing on the chromosphere and transition region.

The paper is organized as follows: In \autoref{Observational data and analysis}, we present the observational data and the pre-processing techniques used in this study. In \autoref{results} we describe the data analysis and the main results, while in \autoref{discussion} and \autoref{conclusion} we discuss and summarize our main conclusions.

\section{Observational data and analysis}\label{Observational data and analysis}

\subsection{Observations}\label{observations}

The observation analyzed here took place on October 9, 2013, within a low-latitude coronal hole (CH). We used multiple IRIS datasets, including slitjaw images (SJI) at 1330~\AA, 1400~\AA, and 2796~\AA\ as well as a very dense raster scan of 3.5~hours in duration and 400 slit positions. 
The field of view was $[140.5, 181.8]$\arcsec, with its center located at $[X,Y] = [435.2, 302.6]$\arcsec\ from the center of the solar disk, and the observation covered the time interval from 16:56~UT to 20:26~UT. We used level-2 data that have been corrected for instrumental effects such as flat-field variations and dark current, and have been wavelength calibrated using the chromospheric O~\textsc{i} 1355.6~\AA\ line.
IRIS provides very high spatial (0.166\arcsec -- 0.35\arcsec) and spectral (40 m\AA) resolution data in the far-ultraviolet (FUV; 1332~\AA\ -- 1407~\AA).

Moreover, we used extreme-ultraviolet (EUV) data and LOS magnetograms, as recorded by AIA and HMI, respectively.
The AIA data have a cadence of 12 seconds and a pixel size of 0.6\arcsec. The 304~\AA\ channel corresponds to the emission of He~\textsc{ii} at a temperature around 7$\times$10$^{4}$ $\mathrm{K}$, depending on the dominant physical process \citep{Antolin_2024}. The HMI data have a pixel size of 0.5\arcsec\ and a cadence of 45 seconds or 720 seconds (12 minutes), with an upper bound of the random noise equal to 10 Mx cm$^{-2}$ or 6.3 Mx cm$^{-2}$ \citep{liu_2012}, respectively.

\subsection{Data treatment}

The 
near-ultraviolet and FUV channels were co-aligned by comparing the fiducial lines of each detector.  
We also performed
radiometric calibration to transform the data numbers into physical units (s$^{-1}$ cm$^{-2}$ sr$^{-1}$  \AA$^{-1}$)\footnote{\url{https://iris.lmsal.com/itn38/analysis_lines_iris.html\#radiometric-calibration}}.
All SDO images were co-aligned with IRIS with the same process as in Paper I. 
For visualization purposes only, we took the average of every three consecutive 45s magnetograms, in order to enhance the signal. Moreover, all HMI images were enhanced with unsharp masking with a 6\arcsec$\,\times$\,6\arcsec\ kernel. The contrast of AIA images was improved with a mild gamma correction ($\gamma = 1.1$).
 
\subsection{Fitting of spectral profiles} 

The fitting process of both the Mg~\textsc{ii} h\&k and Si~\textsc{iv} lines is based on a genetic algorithm (GA), which is an optimization technique that mimics natural selection so that solutions evolve over time to become more optimal \citep{Holland_1992, DEAP_JMLR2012}. Moreover, the profiles of both spectral lines were smoothed before fitting using the Savitzky-Golay filter, which can reduce noise while retaining the overall shape of the profile \citep{Savitzky_Golay}.

\subsubsection{Mg~\textsc{ii} h\&k} \label{fitting of mg profiles}

We fit the profiles with seven Gaussian curves: \footnote{\url{https://iris.lmsal.com/itn39/Mg_diagnostics.html\#the-mg-ii-h-k-lines}} two curves for the k line, two for the h line, and three curves for the blue wing of the k line, the red wing of the h line, and the middle "bump" between the k and h lines, which is formed by the red and blue wings of the k and h lines, respectively. The background is not explicitly treated, but is included in the wings and "bump."

\begin{figure}[hbt!]
 \centering
\resizebox{\hsize}{!}{\includegraphics{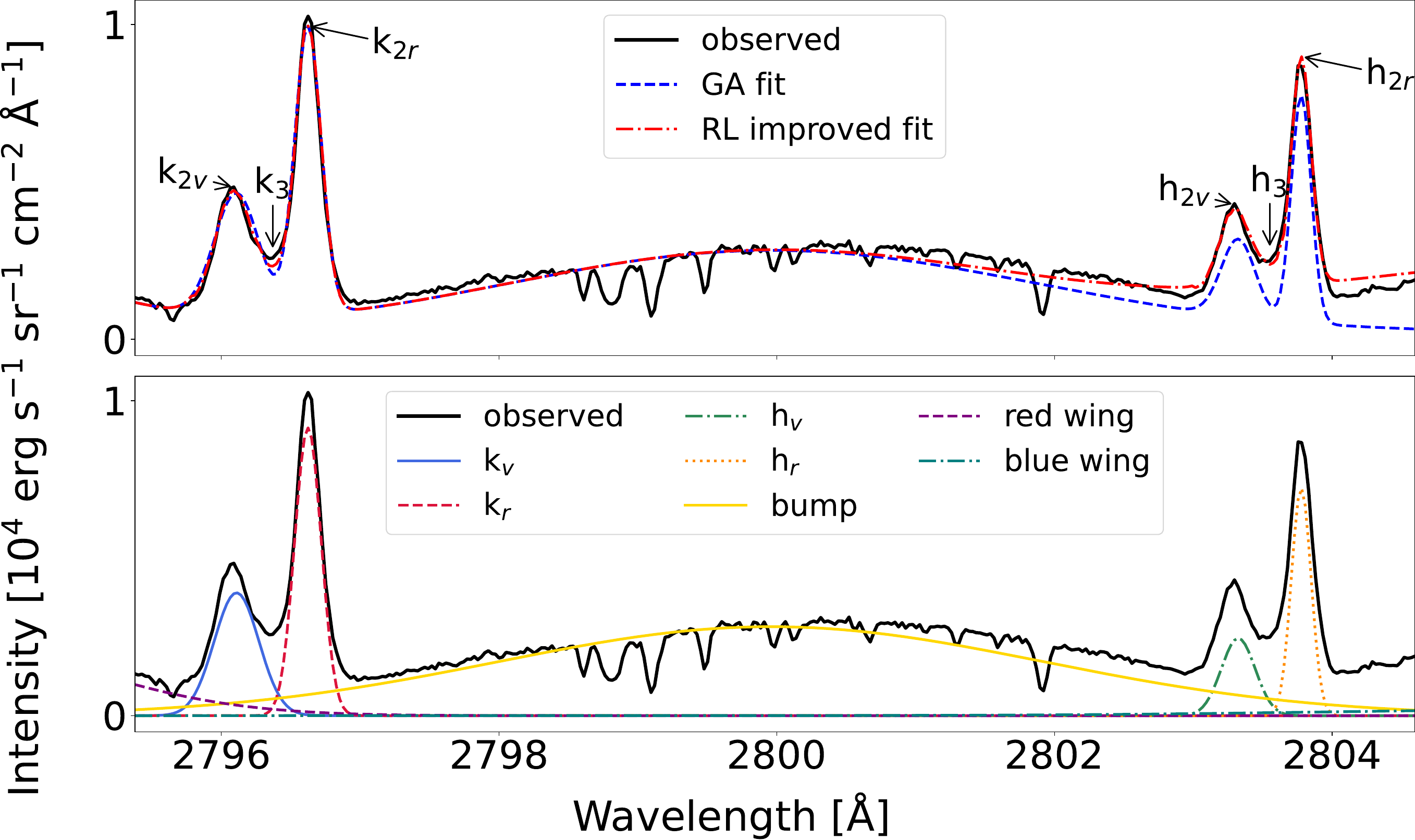}}
\caption{
Example of a typical Mg\,\textsc{ii}\,h\&k spectrum. Top: GA fit and RL improved fit.
The main spectral features are indicated by arrows. Bottom: 
Observed profile with the individual Gaussian components overplotted.
}
  \label{spectral_features}
\end{figure}

In this study, we focus on the Mg~\textsc{ii} k line, with the main spectral features denoted using the standard notation: the line core is referred to as k$_3$, the violet and red emission peaks as k$_{2v}$ and k$_{2r}$, respectively. The features are shown in \autoref{spectral_features} top panel. Although only the k line is analyzed, the simultaneous fit of both lines ensures a consistent representation and precise estimation of the k-line parameters.

The fitting pipeline is the following: First, we applied the least-squares (LS) method, using its parameters as initial values for the GA model, which optimizes each profile fit individually based on the highest coefficient of determination (R$^2$) score \citep{wright1921correlation}, assigning higher weights to the k$_3$ core and the k$_2$ peaks to emphasize the importance of these features. 
The seven-Gaussian fitting helps reproduce the overall shape of the profile, distinguish the different components and reduce the effects of noise and secondary absorption lines, such as the Mn~\textsc{I} 2795.64~\AA\footnote{\url{https://iris.lmsal.com/itn39/photospheric.html}}, which affects the blue wing of the k line. This method is also not restrictive for single-peaked profiles.

However, the algorithm is unable to simultaneously fit the k$_2$ peaks and the k$_3$ dip for profiles with deeper central reversals.
To address this, we utilized a reinforcement learning (RL) algorithm, based on a Deep Q-Network \citep{Mnih_2013_dqn}. The RL agent was applied to the seven-Gaussian fit, iteratively adjusting the values around the peaks and central reversals. By maximizing the weighted coefficient of determination R$^2$ between the fitted curve and the observed profile, the final fit was improved.

The bottom panel of \autoref{spectral_features} shows an example of a fitted profile. The fitting performance is significantly improved at each stage. Using only the LS method, 45$\%$ and 39$\%$ of the profiles have R$^{2} < 0.85$ and R$^{2} > 0.95$, respectively. These values improve to 8$\%$ and 47$\%$  with the GA optimization, and further to 6$\%$ and 58$\%$ with the RL adjustment. 
We used the RL fits in the following analysis, excluding profiles with $R^{2}<0.85$. 
For double-peaked profiles, the k$_3$ position is determined as the minimum between the two maxima that represent the k$_2$ peaks.

\subsubsection{Si~\textsc{iv} 1393.75~\AA} \label{si iv fitting}

In Paper I, we presented the fitting of asymmetric profiles with a bimodal Gaussian. Here we improve this process by first fitting each Si~\textsc{iv} line with both a single and a bimodal Gaussian curve, and then using a GA to optimize the parameters of the bimodal fit. The best model is selected based on multiple statistical criteria, including the Likelihood Ratio Test \citep{wilks1938lrt}, the Akaike Information Criterion \citep{akaike1974aic}, the Bayesian Information Criterion \citep{schwarz1978bic}, and Bayes factors \citep{kass1995bayes}. If both lines strongly support a bimodal fit and the two fitted curves exhibit similarity constraints, the bimodal model is chosen; otherwise, the single Gaussian model is preferred.

\section{Results} \label{results}

The section is organized into two parts. The first focuses on the evolution of a dynamic absorption feature in AIA 304~\AA, probably a mini-filament, leading to the jet formation, as well as the analysis of the magnetic field and chromospheric diagnostics. 
The second part presents a statistical analysis of the Mg~\textsc{ii} k and Si~\textsc{iv} 1393~\AA\ diagnostic parameters for the key structures discussed in Paper I.

In Paper I, we analyzed the jet evolution across multiple AIA channels. The jet first appears in the AIA 304~\AA\ channel and subsequently in higher-temperature channels. The cool and hot counterparts of the jet spire do not spatially coincide with each other. A loop appears simultaneously, with its southern footpoint rooted at the same bright point from which the jet originates. A plume, rooted at the edges of the CH, acts as foreground emission. 
The jet and the loop are visible in most AIA channels, as well as the Si \textsc{iv} spectra and 1400~\AA\ SJI. 
Here, we also report a new feature: a bright redshifted structure that is clearly visible in the chromosphere but absent in higher temperatures.

\subsection{Pre-eruption evolution and jet dynamics} \label{mini-filament} 

\subsubsection{Evolution leading to the jet eruption} \label{images} 

\begin{figure*}[hbt!]
    \centering
     \centering\resizebox{\hsize}{!}{\includegraphics[width=\linewidth]{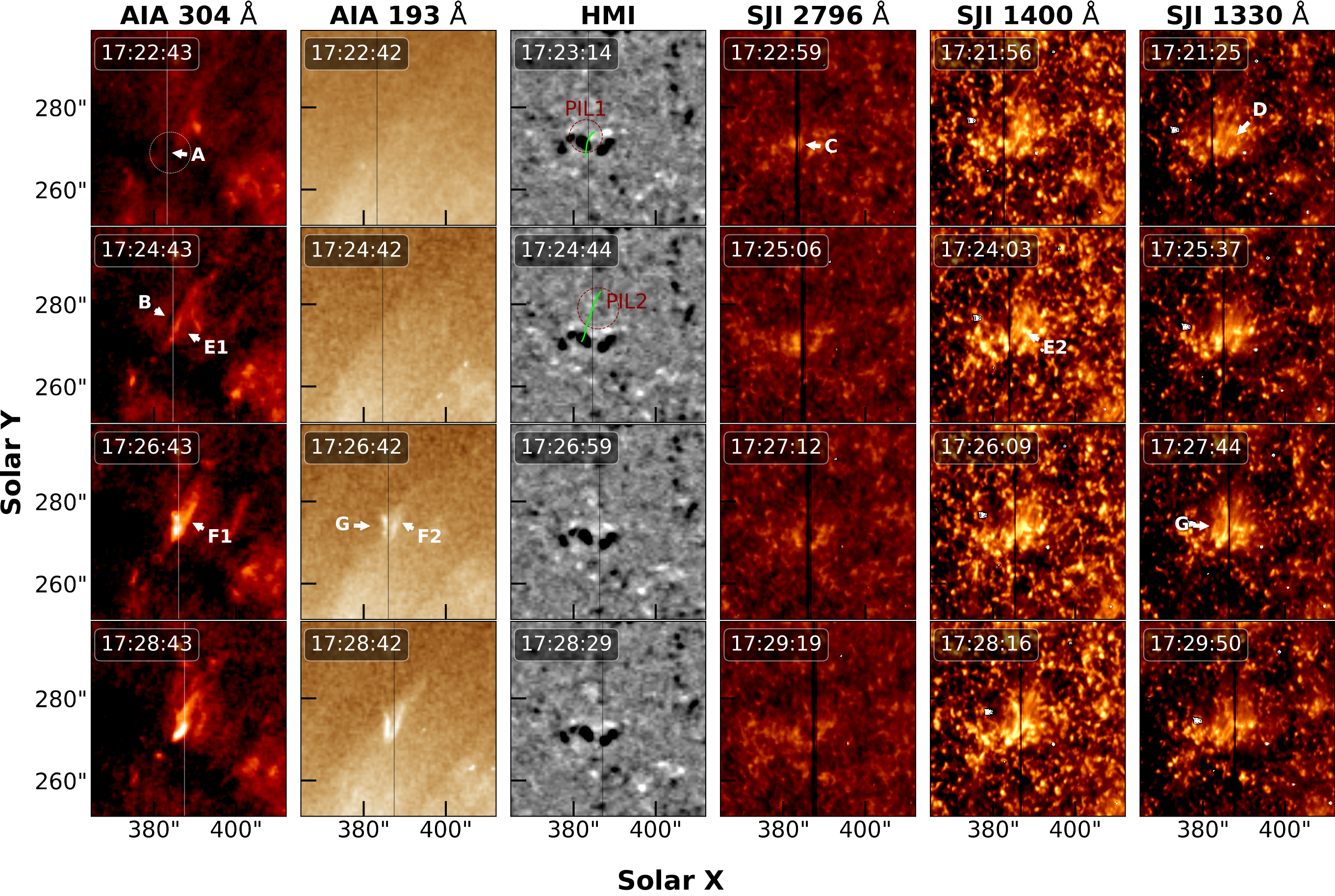}}
    \caption{Temporal evolution of the jet as observed by multiple instruments.  
  The green lines denote the approximate projected mini-filament-like structure tracing, as observed on AIA 304~\AA. The dashed red circles at 17:23:14 and 17:24:44 show the two PILs (see text). The vertical lines on AIA 304~\AA\ and 193~\AA\ columns indicate the position of the IRIS slit. 
Arrows: A, the initial position of the mini-filament-like structure (also enclosed by the dashed white circle); B, the elongated mini-filament-like structure; C, the bright redshifted structure; D, the pre-existing bright features; E1, the elongated feature; E2, structures possibly below the elongated feature; F1, the cool jet component; F2, the hot jet component; G, the loop associated with the jet eruption. \\ Online animations: AIA 304~\AA\ and AIA 193~\AA, with arrows annotating selected structures.
}
   \label{all images}
\end{figure*}

Prior to the jet eruption, we observe a well-defined dark elongated absorption feature (\autoref{all images} arrow B).
Before that, it is close to the resolution limit of the instruments and thus difficult to distinguish from its surroundings.
We believe that this structure is most likely a mini-filament, as it appears in absorption in AIA 304~\AA, but is invisible in the 1400~\AA\ and 1330~\AA\ SJ images, indicating a cool structure containing neutral hydrogen, neutral and singly-ionized helium absorbing photons of $\lambda$~<~912~\AA\ \citep{Williams_2013,Anzer_2005}.  Note that it is not clearly visible in AIA 193~\AA\ and AIA 171~\AA, due to the plume emission. \cite{Adams_2014} report on a similar absorption structure.

The green lines in the HMI panels at 17:23:14 and 17:24:44 UT show the exact position of this structure, as projected in AIA 304~\AA\ at similar timestamps. 
Until 17:23:43 UT, it appears as a small, dark structure roughly above a small polarity inversion line (PIL1) between a weak positive and a strong negative magnetic field patch. Its length can be roughly estimated up to approximately 3.5 Mm (see \autoref{all images}, arrow A and online animation). It possibly forms at approximately 17:17 UT; however, it is very faint and can hardly be distinguished from the background until 17:23:43 UT, making both its exact formation time and its identification as a mini-filament uncertain. For this reason, we refer to it as "mini-filament-like structure" hereafter.

\begin{figure}[hbt!]
    \centering
     \resizebox{\hsize}{!}{\includegraphics{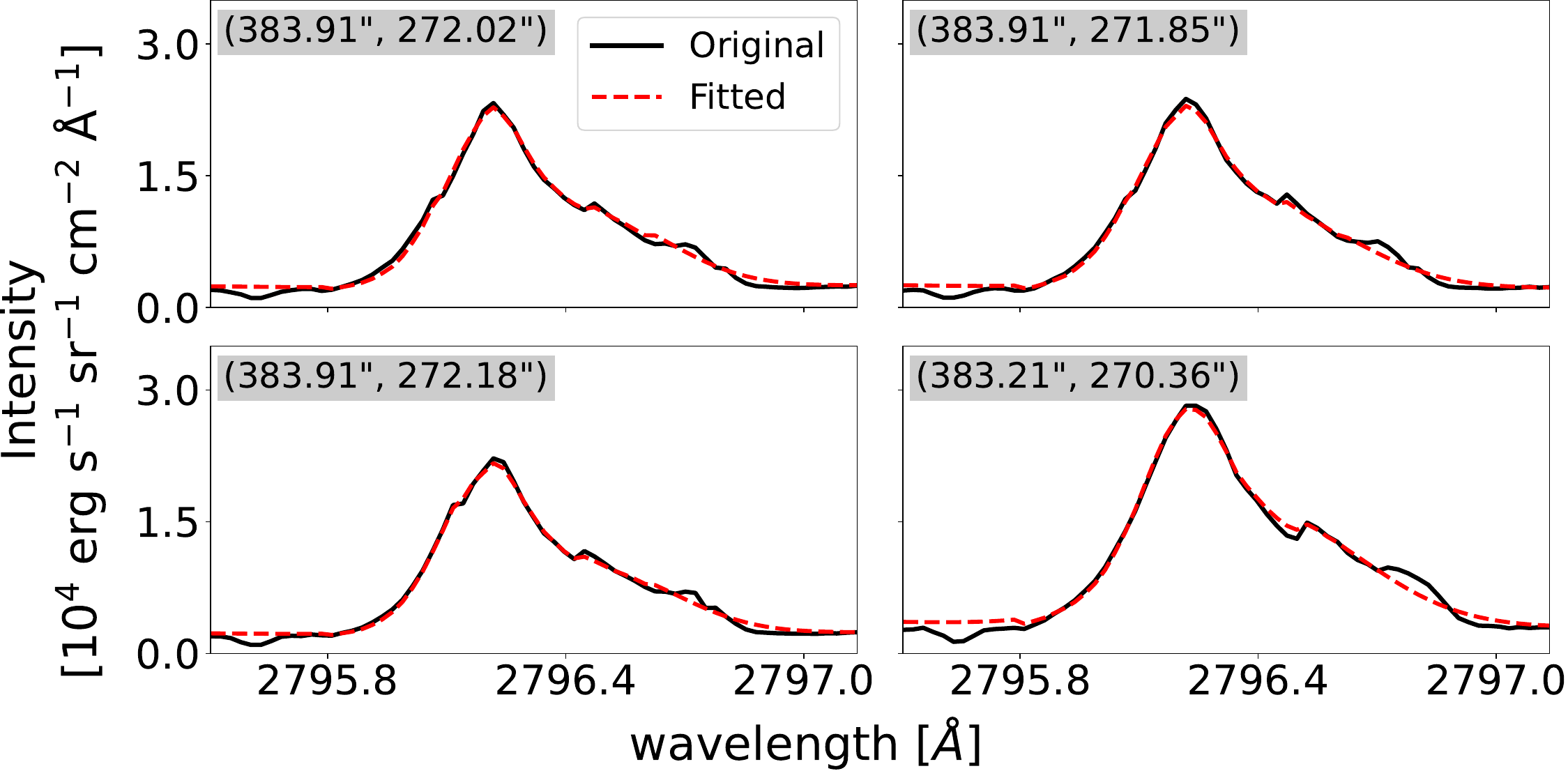}}
    \caption{Four single-peaked Mg~\textsc{ii} k profiles on the bright redshifted structure. The dashed red line is the fitted seven-Gaussian curve, and the black line the actual profile (see paragraph~\ref{velocities}).}
    \label{random_cool_profiles}
\end{figure}

A bright structure is observed at the same location at SJI 2796~\AA\ (arrow C at 17:22:59 UT, \autoref{all images}), also captured by the IRIS slit from 17:20:53 to 17:23:31 UT, evident in the Mg~\textsc{ii} k raster at (382\arcsec, 270\arcsec), close to the southern part of the loop and the jet base (see also \ref{velocities}). The Mg~\textsc{ii} profiles on this structure are almost or completely single-peaked. An example of four such profiles is shown in \autoref{random_cool_profiles}. Hereafter we refer to it as "bright redshifted structure" and its exact nature is analyzed in detail in \autoref{discussion}.

Following again the evolution of the mini-filament-like structure, it exhibits upward motions, appearing more compact and increasing in length from 17:23:43 to 17:24:43 UT. Mini-filaments are observed to increase in length \citep{Panesar_2017}, as well as large-scale filaments \citep{Karki_2026}. During this evolution it reaches a length of 11 Mm (15\arcsec) and its southern end is still located close to PIL1, while below its middle section a second, weaker PIL (PIL2) is also found (red circle at 17:24:44).

Simultaneously with the movement of the mini-filament-like structure, from 17:23:43 to 17:24:43 UT, we observe an elongated feature, visible in AIA 304~\AA\ (arrow E1 on \autoref{all images}). It forms concurrently with the mini-filament-like structure, spanning from (384\arcsec, 270\arcsec) to (395\arcsec, 285\arcsec). 
The high spatial resolution of the SJ 1400~\AA\ images (arrow E2) suggests that the elongated feature possibly consists of multiple bright fine-scale structures unresolved by AIA or, alternatively, that these structures lie below the mini-filament-like structure and are thus only partially visible.

At 17:24:55 UT, the southern end of the elongated feature (385\arcsec, 270\arcsec), also close to the southern end of the mini-filament-like structure, is intensified in AIA 304~\AA\ (see animation). At the same time, the mini-filament-like structure continues its rising motions, until it disappears. Shortly thereafter, the cooler jet component (arrow F1) appears at 17:25:19 UT (see animation), followed approximately 1 to 2 minutes later by the hotter jet component (arrow F2) and the loop (arrow G in \autoref{all images}), both originating from the same location \citep{Sterling_2015}.

\subsubsection{Magnetic field evolution}

The HMI 45\,s magnetograms show no significant evolution of the photospheric magnetic field minutes before the jet eruption (see Paper I). 
To further assess temporal variations in magnetic field strength, we used HMI 720\,s magnetograms up to approximately 2 hours before the jet eruption. We computed the summed magnetic field over a fixed subregion around PIL1 beneath the mini-filament-like structure ($\sim$200 pixels), treating the positive and negative polarities separately. We also calculated the net unsigned flux, which is the sum of the positive and absolute negative flux. To avoid noise contamination, we selected pixels with values greater than the nominal noise level of 6.3 Mx cm$^{-2}$ \citep{liu_2012}. The selected region was also kept fixed and fully enclosed the magnetic polarities of interest, minimizing variability due to magnetic elements moving into or out of the box.

\begin{figure}[hbt!]
    \centering
    \begin{subfigure}[c]{0.99\linewidth}
        \centering 
        \includegraphics[width=\linewidth]{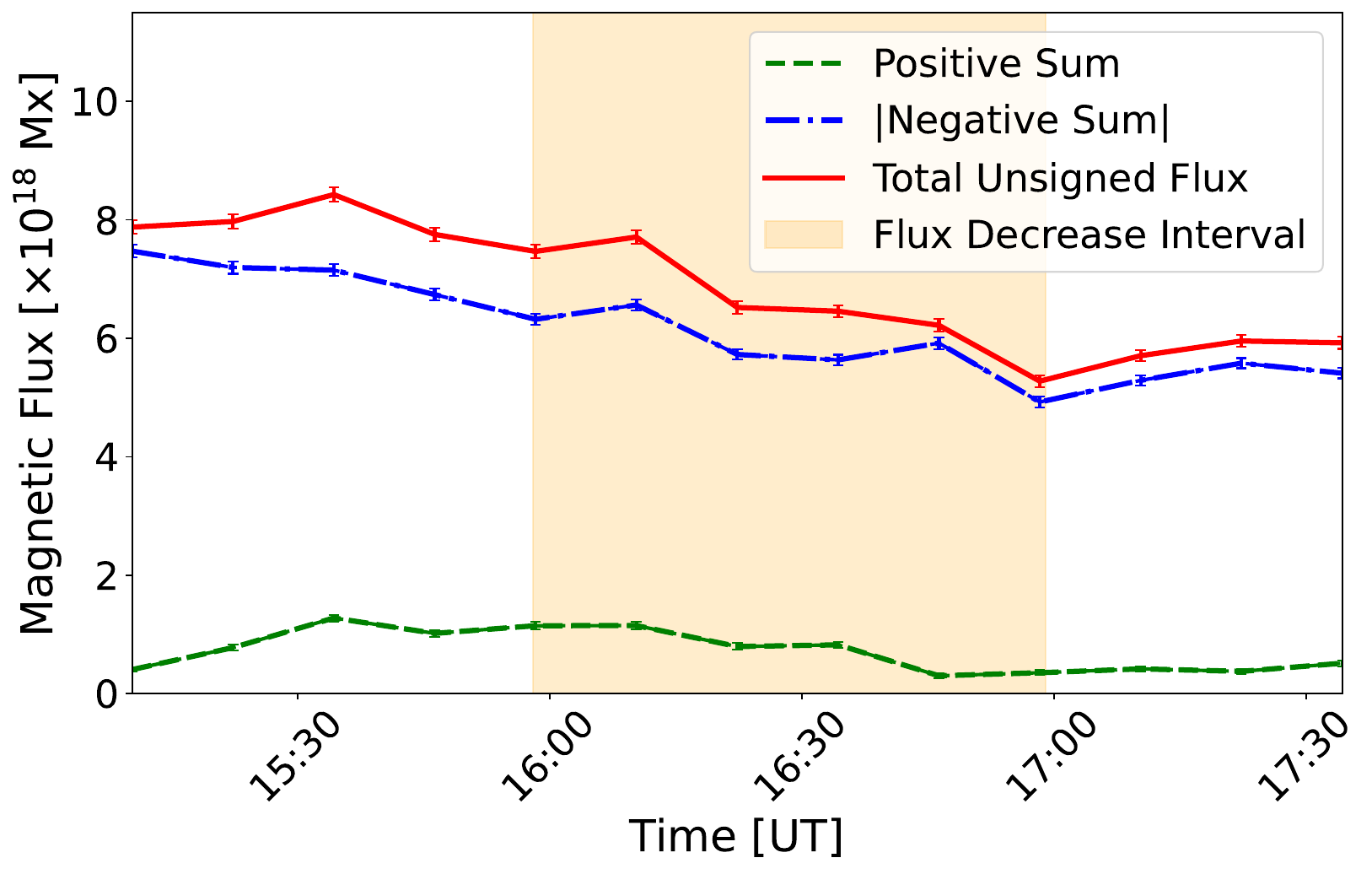}
    \end{subfigure}

    \begin{subfigure}[c]{0.99\linewidth}
    \includegraphics[width=\linewidth]{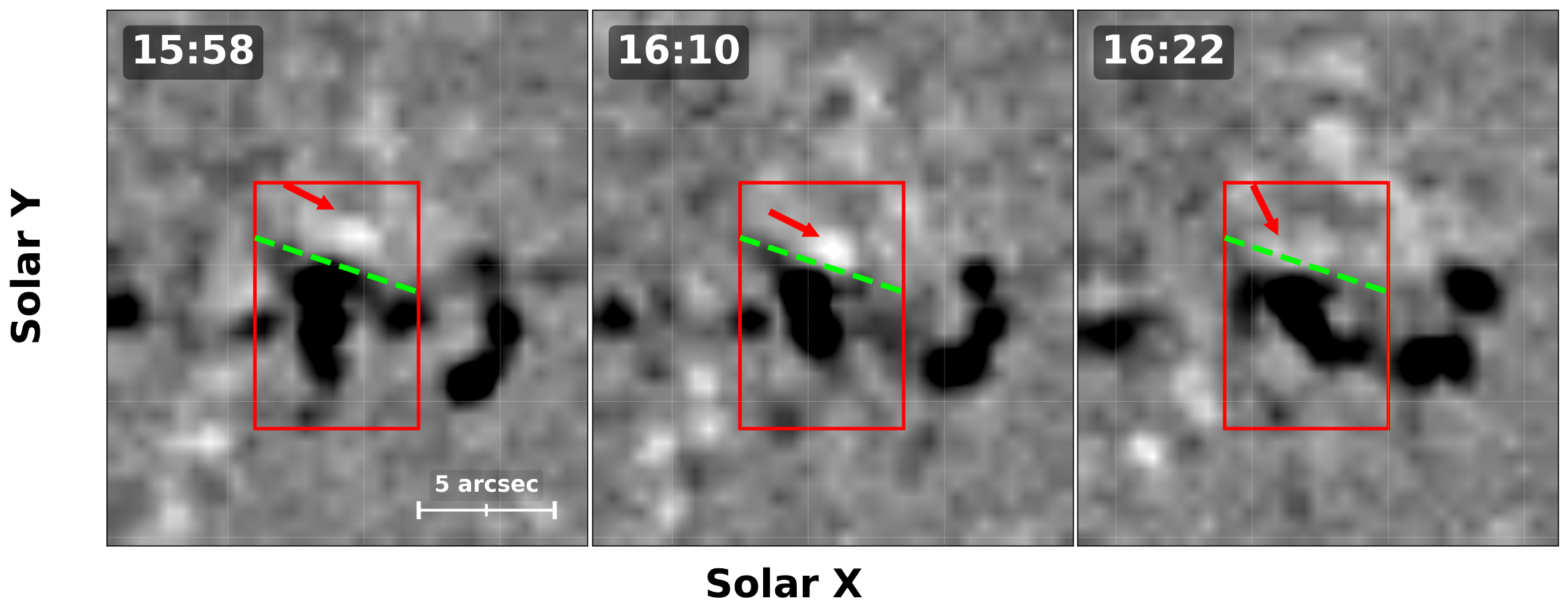}
    \end{subfigure}

\caption{Top: Temporal evolution of the integrated HMI magnetic flux with error bars. The orange shaded area shows the duration in which we observed a rate of decrease of 10$^{18}$ Mx/hour. Bottom: Three HMI 720\,s magnetograms showing the red box within which we calculated the aforementioned fluxes. The dashed green line indicates PIL1. The red arrow shows the area of decrease and shear motion of the positive polarity.
\\
An online animation of the HMI 720\,s magnetograms is available.}

    \label{fig:hmi 720}
\end{figure}

The top panels of \autoref{fig:hmi 720} show the temporal variation of the positive, negative and unsigned fluxes. 
Note that positive and negative fluxes are not balanced, because the dominant polarity of the CH is negative. 
The bottom panels show the area within which these values are calculated with a red box in two different timestamps. The green dashed line marks PIL1.

From 15:10 UT to 15:46 UT we observe flux increase within and around the red box, which may indicate magnetic flux emergence. 
At 15:46 UT a positive polarity starts to move toward PIL1 and the negative polarity, leading to decrease of both polarities from 15:58 UT to 16:58 UT (orange shaded area on the top panels of \autoref{fig:hmi 720} and red arrow at 15:58 and 16:10 UT). We also observe motions of positive patches almost parallel to PIL1, possibly suggesting shear motions (red arrow 16:22 UT \autoref{fig:hmi 720}). Both phenomena can also be seen in the online HMI animation. The total decrease rate of the total unsigned flux is $2.2 \times 10^{18}$~Mx h$^{-1}$, which is comparable to the observed values reported by \cite{Chen_2025_mf} and simulations \citep{Patsourakos_2025} but slightly larger than those from \cite{Panesar_2018}. For the positive flux the rate is  $0.8 \times 10^{18}$~Mx h$^{-1}$ and the negative is  $1.4 \times 10^{18}$~Mx h$^{-1}$. 
The percentage decrease is 70$\%$ for the positive and 20$\%$ for the negative flux, while for the total unsigned flux is 30$\%$. From 16:49 to 17:11 UT, recurrent increases in intensity are observed for several minutes at the same location (see AIA 304~\AA\ animation).

Given that the upper bound of the noise level for the 720\,s magnetograms is \(\mathrm{\sigma_B=6.3\ Mx\ cm^{-2}}\) and assuming that all pixels are independent with each other, the total error of the summed flux is of the order of
$10^{17}$~Mx. The corresponding errors for the positive and negative fluxes are of the order of $10^{16}$~Mx. All errors are smaller than the measured flux by at least one order of magnitude, indicating that the measured decrease is significant. However, the above values are actually lower limits, as in reality there is some level of correlation between pixels. 

\subsubsection{Magnetic field modeling}

We study the magnetic topology of the observed structures by applying a potential magnetic field extrapolation to the 720\,s magnetogram obtained at 17:22:16 UT. We used a method\footnote{\url{https://dev.flarecast.eu/stash/projects/FE/repos/nn_currents-idl/browse/lff_extrap.pro}} based on the Fast Fourier Transform approach of \cite{Alissandrakis_1981}. To minimize boundary effects, the extrapolation was initially applied to a region surrounding the CH that extends twice the field of view in both $X$ and $Y$ directions. This way the area below the jet lies approximately at the center of the extrapolation.

\begin{figure}[hbt!]
    \centering
    \begin{subfigure}[c]{0.45\linewidth}
        \centering      \includegraphics[width=\linewidth]{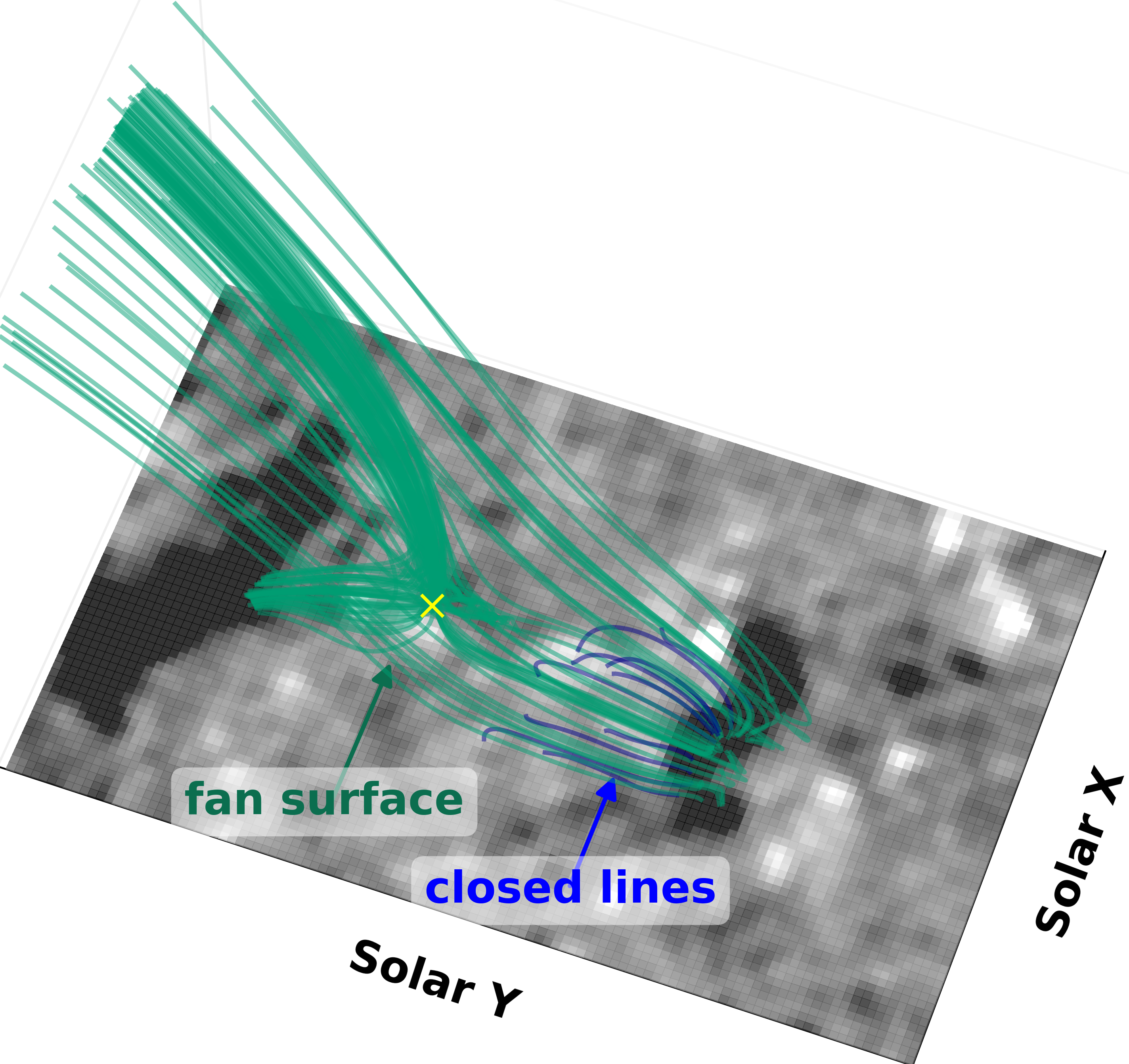}
    \end{subfigure}
    \begin{subfigure}[c]{0.54\linewidth}
        \centering
    \includegraphics[width=\linewidth]{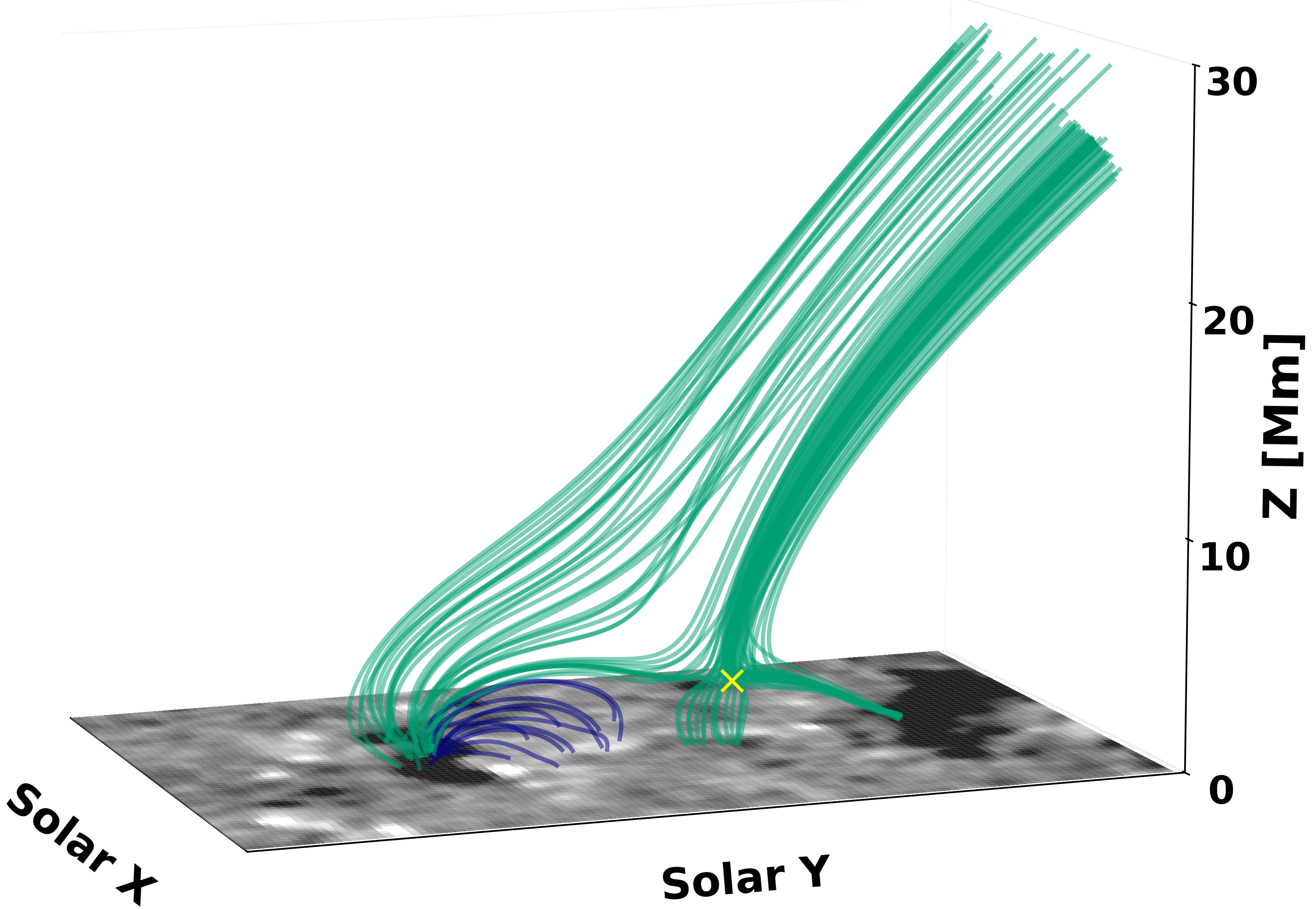}
    \end{subfigure}
    
    \begin{subfigure}[c]{0.32\linewidth}
        \centering
        \includegraphics[width=\linewidth]{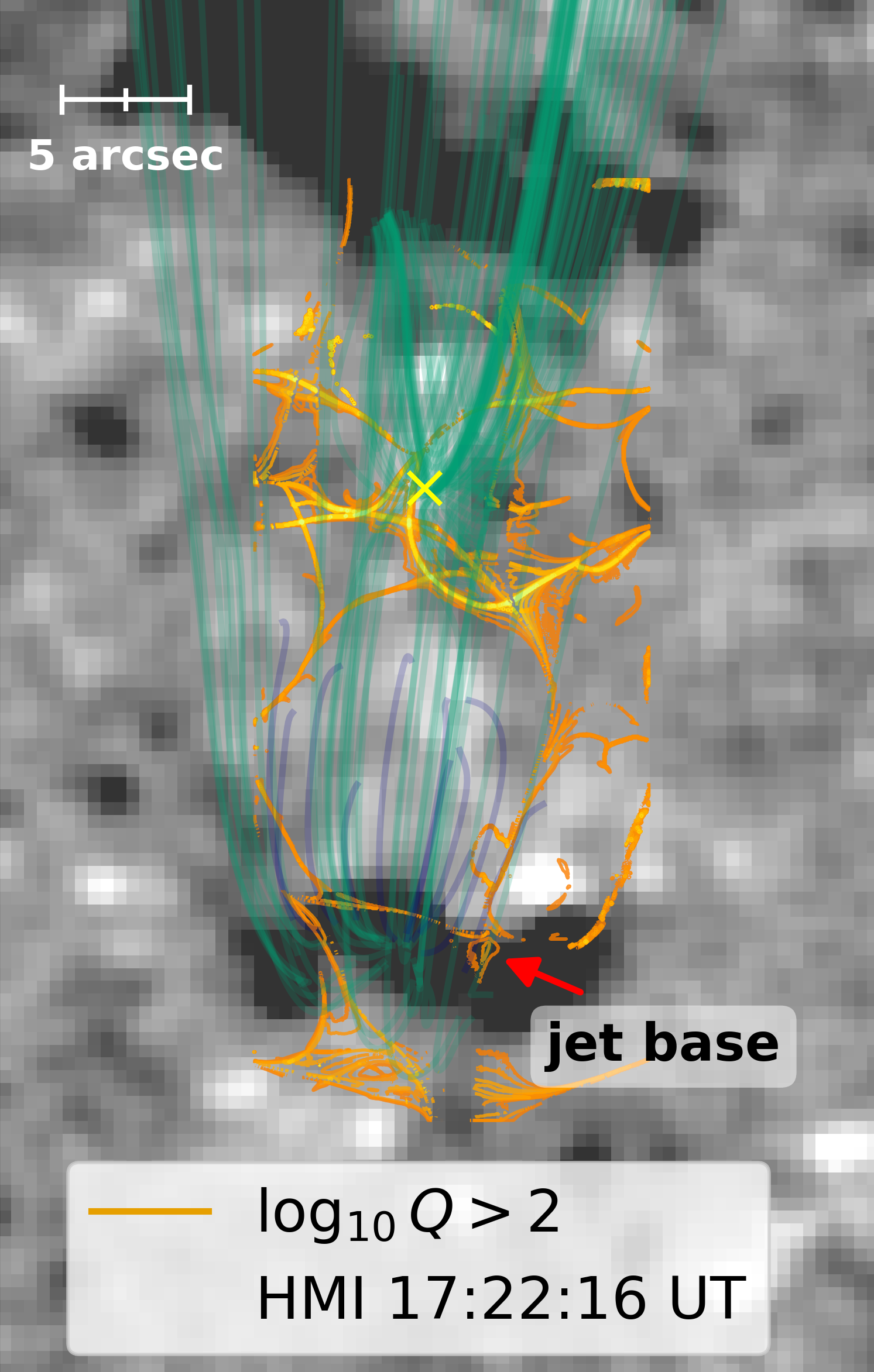}
    \end{subfigure}
    \begin{subfigure}[c]{0.32\linewidth}
        \centering
        \includegraphics[width=\linewidth]{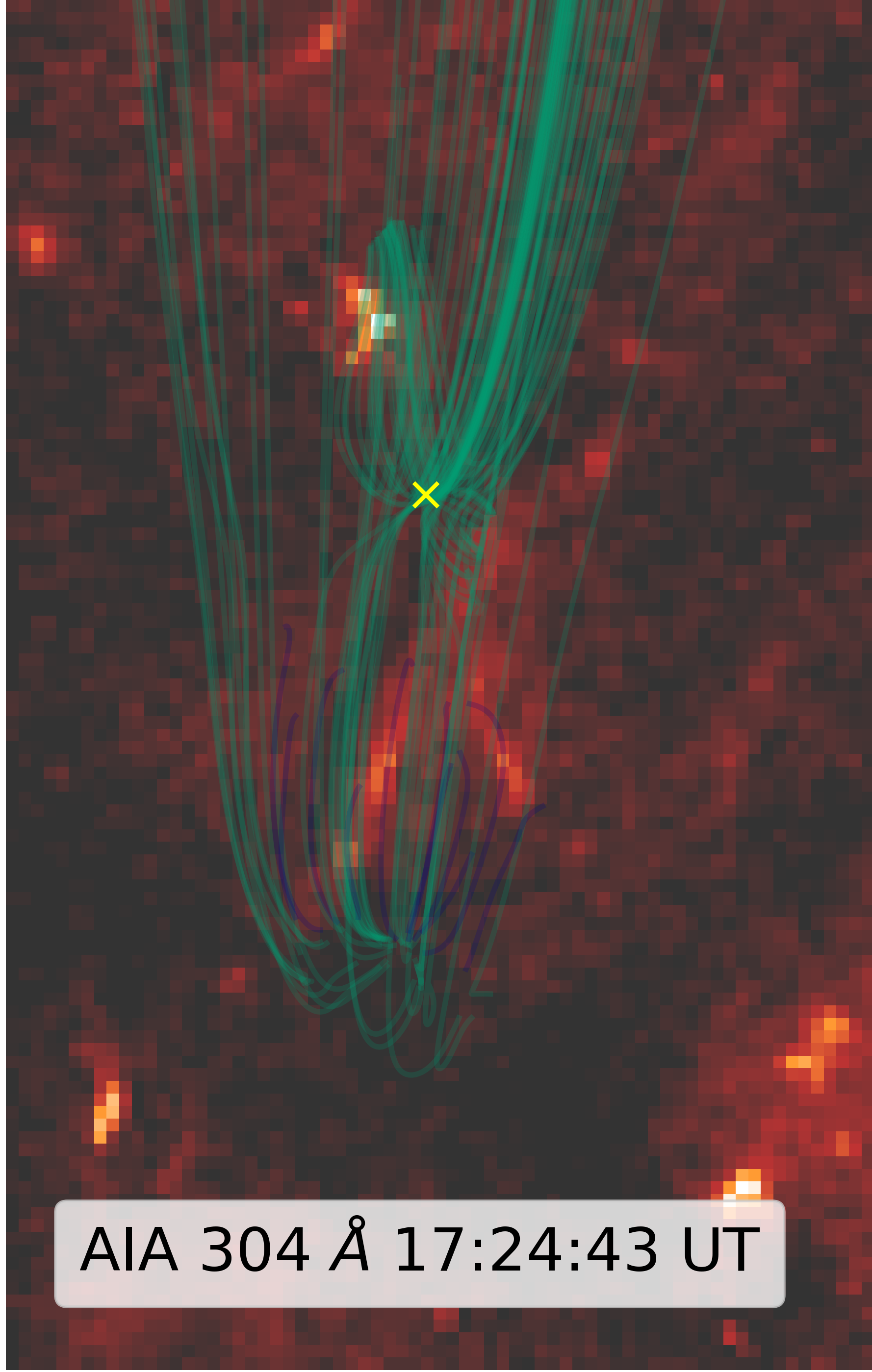}
    \end{subfigure}
    \begin{subfigure}[c]{0.32\linewidth}
        \centering
        \includegraphics[width=\linewidth]{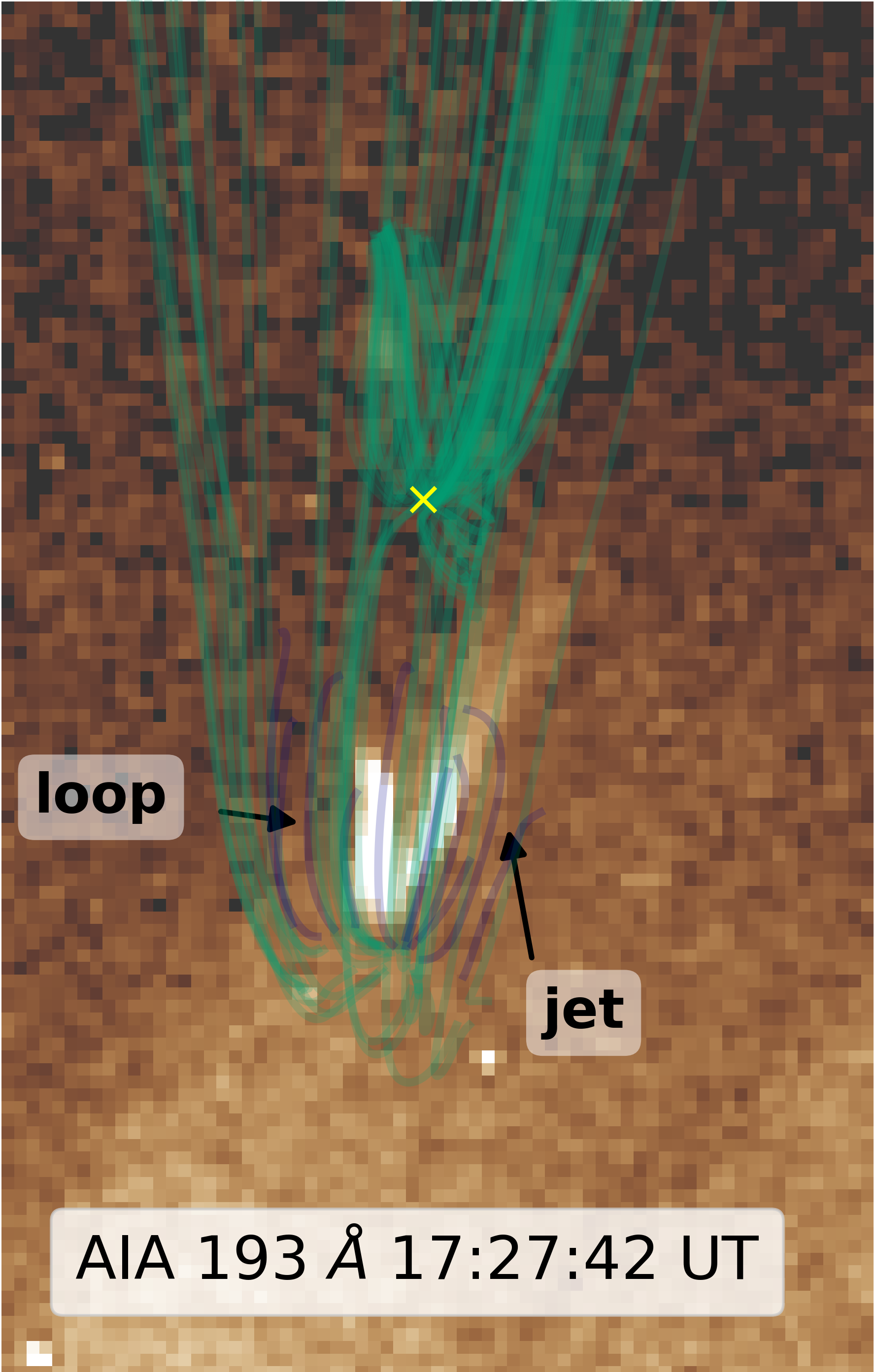}
    \end{subfigure}
    \caption{Results of the potential field extrapolation. The green lines correspond to open magnetic field lines, while the blue lines correspond to closed ones. Top row: View from different angles over the extrapolated B$_z$ at z=0. Bottom row, from left to right: View from 90$^\circ$ elevation angle over the extrapolated B$_z$ at z=0, AIA 304~\AA, and 193~\AA. The displayed Q values are calculated at 2 Mm, with higher values shown in yellow. The yellow star denotes the null point, while green and blue lines are open and closed magnetic field lines, respectively.}
    \label{fig:hmi 720_extrapolation}
\end{figure}

\autoref{fig:hmi 720_extrapolation} shows the results of the above calculations. The topology provides useful insight into the magnetic connectivity of the region. We observe closed magnetic field lines of length up to 12\arcsec, spatially coincident with the loop (\autoref{all images} arrow G) and open lines aligned with the jet spire.

To further investigate the magnetic topology, we searched for magnetic null points, using the Null$\_$Finder module \footnote{\url{https://github.com/FedericaChiti/Null_Finder/blob/main/null_finder.py}}, which implements a trilinear interpolation and the Newton-Raphson root-finding algorithm \citep{Haynes_2007}. 
Moreover, we calculated the squashing factor $Q$ using the FastQSL code \citep{Titov_2002,Zhang_2022}, which employs a GPU-accelerated adaptive Runge--Kutta--Fehlberg (RKF45) integration scheme.

We found a null point at a height of 2 Mm above a photospheric parasitic (positive) polarity, also close to where the northern edge of the mini‑filament-like structure is located 2.5 minutes later (\autoref{fig:hmi 720_extrapolation}, AIA 304~\AA, 17:24:43 UT). 
Around the null point, a small fan-like topology is evident, anchored in surrounding negative magnetic field polarities. 
The map of $\log_{10}$Q displayed in \autoref{fig:hmi 720_extrapolation} corresponds also to 2 Mm height.  High $Q$ values, $\log_{10}$Q$\gg$2 \citep{Titov_2002}, indicate the presence of QSLs, where reconnection is likely to occur. The null point lies near the intersection of Q structures, indicating strong changes in magnetic connectivity in its vicinity.
The jet originates at the southern end of the fan surface, where regions of $\log_{10}$Q>2 separate the positive from the negative magnetic polarities (red arrow in \autoref{fig:hmi 720_extrapolation}).

The apparent misalignment between the jet and the outer spine of the fan may be attributed to projection effects. Another possibility is that the potential field extrapolation may not accurately reproduce the precise location of the null point, since mini-filaments carry electric currents that can shift its position \citep{Regnier_2012}. Nonetheless, the overall magnetic topology indicates that reconnection likely plays a key role in the activation of the mini‑filament-like structure and the subsequent jet eruption.

\subsubsection{Inversions of IRIS observations} \label{inversions} 

To retrieve the optical-depth-dependent atmospheric properties of the jet region, we employed the IRIS$^2$ inversion code \citep{Sainz_Dalda_2019}, based on the STockholm Inversion Code (STiC) framework \citep{De_la_Cruz_2019}. The inversion is performed with a node spacing of $\Delta \log \tau = 0.2$, using the Mg \textsc{ii} h\&k lines and the Mg \textsc{ii} UV triplet\footnote{\url{https://iris.lmsal.com/iris2/iris2_chapter01.html}}.
STiC implements non-local thermodynamic equilibrium radiative transfer with partial frequency redistribution and node-based parametrization to produce a smooth stratification of temperature, velocity, density and microturbulence. These capabilities make it particularly well suited for diagnosing the dynamic chromosphere.

\begin{figure}[hbt!]
  \centering\resizebox{\hsize}{!}{\includegraphics{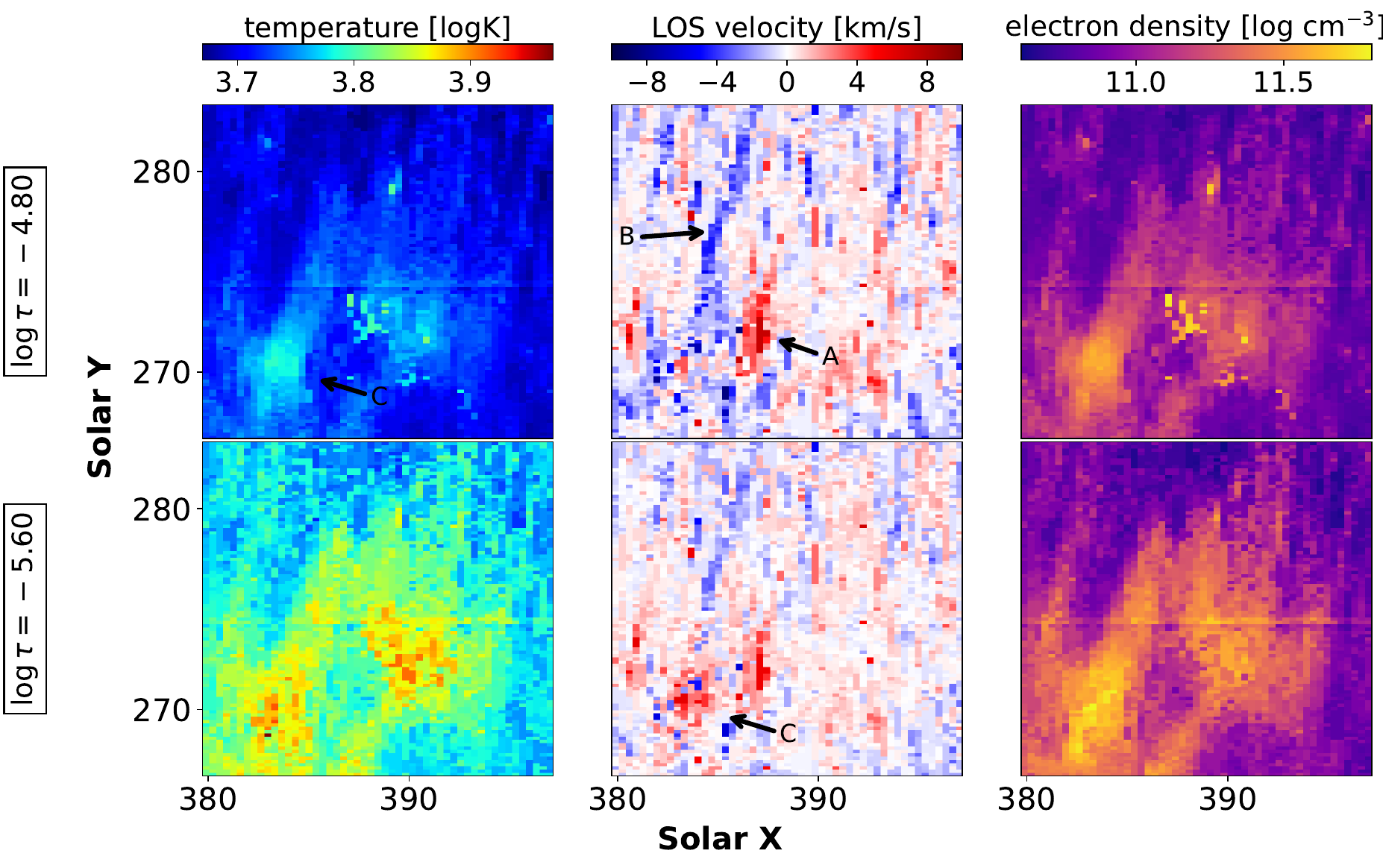}}
  \caption{Inversion maps of the jet region. Each row corresponds to a different $\log\tau$ value, indicated at the left of each row. Arrows: A, jet base; B, blueshifted stripe; and C, bright redshifted structure.}
  \label{inversion_maps}
\end{figure}

Inversions provide constraints on the variation of plasma parameters with optical depth \citep{Sainz_Dalda_2019}, beyond the profile fitting methods presented in paragraph~\ref{fitting of mg profiles}. The range
$-6.5<\log\tau<-5$ is representative of the middle chromosphere and $-5<\log\tau<-4$ of the low chromosphere \citep{Sainz_Dalda_2023}. The most reliable values of the model are those for $\log\tau = -5$, where $\tau$ is calculated at 5000~\AA\ \citep{Sainz_Dalda_2019}. 
For this reason, we select $\log\tau = -4.8$. We also decide to include $\log\tau = -5.6$, as it is the closest optical depth where the bright redshifted structure is evident and is relatively close to $-5$. However, these results should be interpreted with caution.  
Our findings are shown in \autoref{inversion_maps}.  

At $\log\tau = -4.8$, at the base of the jet (arrow A), we observe redshifted velocities up to 7 km/s, which could indicate plasma flows opposite to the direction of the jet eruption. Systematic redshifts are also present at the jet base at $-4.8 < \log\tau \leq\ -4$. In addition, the blueshifted stripe (arrow B) possibly tracks the upward motion of the mini-filament-like structure with a velocity of 4 km/s. At the same location, we also observe lower temperatures (5000~K) and lower electron densities ($6.5 \times 10^{10}$ cm$^{-3}$). 
At this optical depth, a few profiles exhibit redshifted velocities on the bright redshifted structure, which is also denser ($2.5\times10^{11}$ cm$^{-3}$) and hotter (5800~K) than the surrounding regions. 
As already mentioned, the bright redshifted structure (arrow C) exhibits systematic redshifted velocities of 5~km/s at $\log\tau=-5.6$.

\subsubsection{Velocities and intensities} \label{velocities}

We used the red-blue (RB) asymmetry to assess the plasma motions. 
The RB asymmetry \citep{Leenaarts_2013b,Pereira_2013}, is the normalized difference of symmetrical integrals of the intensity profile around the core, defined as \(\text{RB} = (J_v - J_r)/(J_v + J_r)\), 
where
\(
J_r = \int_{v - \Delta v}^{v + \Delta v} I(v) \, dv
\quad \text{and} \quad
J_v = \int_{-v - \Delta v}^{-v + \Delta v} I(v) \, dv
\)
 represent the integrals over the red and violet sides of the profile, respectively. Here, we set \(v = 25~\text{km/s}\) as the velocity offset around the k$_3$ core, corresponding to the k$_2$ peaks, and \(\Delta v = 10~\text{km/s}\) as the width of the integration window. The integrals were evaluated using the Simpson rule \citep{Cartwright_2017}. 

Thus, the RB asymmetry is a measure of the variation of velocities (along the line of sight) above the peak formation height. In a simple example, suppose that the plasma of the upper chromosphere moves upward while the middle chromosphere is at rest. The photons with wavelength near $\lambda_{k_{2v}}$ will face a higher absorption probability  than the ones with wavelength near $\lambda_{k_{2r}}$. In other words, blueshifted photons emitted from deeper layers are mostly absorbed, leading to a weaker violet $k_2$ peak.
In contrast, a downward velocity enhances the blue $k_{2v}$ peak and weakens the $k_{2r}$ one
(see \cite{Pereira_2013} and \cite{Carlsson_1997} for the same effect in the Ca~\textsc{ii} H$\&$K lines).  

We used the following spatial masks to identify and distinguish the different structures present in the observed region; the jet and loop masks are based on the Si~\textsc{iv} 1393~\AA\ Dopplergrams (see Figure 5 from Paper I), as they correspond to regions of significant blueshifted and redshifted velocities, respectively. Note that the boundaries of these regions cannot be strictly determined, and some uncertainty remains.  We also note that, the "loop mask" refers to the region below the loop in Mg \textsc{ii} diagnostics, as this structure is not visible in the Mg \textsc{ii} k maps. The "jet mask" refers to its cooler component, since this and not the hotter component is visible in the Si~\textsc{iv} spectra.

\begin{figure}
    \centering
 \centering\resizebox{\hsize}{!}{\includegraphics{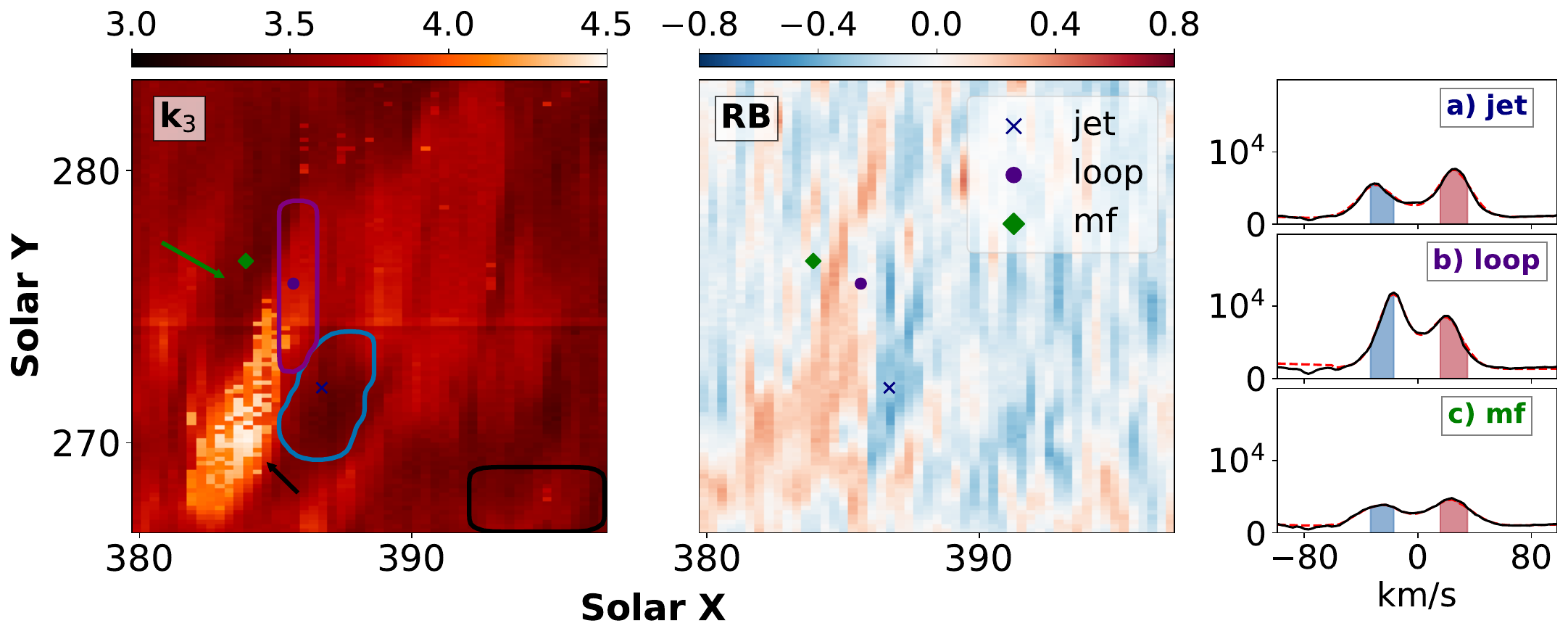}}
\caption{% 
Left: Map of logarithmic Mg \textsc{ii} k$_3$ intensity. Middle: Map of the RB asymmetry of the Mg~\textsc{ii} k line calculated over the velocity intervals 
\([-35, -15]\) and \([15, 35]\)~km/s. 
Right: Characteristic profiles corresponding to each region marked on the maps. 
The solid black line denotes the observed profile, while the dashed red line represents the fitted curve. 
Contours are color coded as follows: blue for the jet spire, purple for the loop, 
and black for the background region. The green and black arrows show the mini-filament-like structure (denoted with "mf") and bright redshifted structure. 
The blue (blueshift) and red (redshift) shaded intervals mark the spectral ranges used to compute the RB asymmetry maps. Intensities are in units of erg s$^{-1}$ sr$^{-1}$ cm$^{-2}$ $\AA^{-1}$.
}
\label{rb maps}
\end{figure}

\autoref{rb maps} shows the logarithmic Mg \textsc{ii} k$_3$ intensity map, with the contours showing the jet spire, loop and background region over-plotted. The arrows point to the bright redshifted structure and the mini-filament-like structure at its ascending phase (see also arrow B on \autoref{inversion_maps}). The RB asymmetry map and example profiles of the jet spire, loop and mini-filament-like structure are also shown. The mini-filament-like structure appears as a darker region, with lower k$_3$ and peak intensities (see example profile). Some profiles on the bright redshifted structure are significantly brighter, as the absence of a central reversal causes the algorithm to identify the intensity maximum as k$_3$. 
The RB asymmetry shows downward motions at the bright redshifted structure in the middle chromosphere, which indicates plasma falling toward the solar surface. This result is consistent with the redshifted velocities at $\log \tau\,=\, -5.6$ derived with IRIS$^2$ (see paragraph~\ref{inversions}). At the same time, blueshifted velocities are detected at the jet spire, suggesting small upflows originating in the chromosphere.

 We do not find systematic k$_3$ Doppler velocities on the jet spire, or any indication of upward motions of the mini-filament, as we did with IRIS$^2$.  However, a small number of profiles on the jet exhibit blueshifted Doppler velocities up to 6 km/s. On the other hand, we find redshifted values  below the loop of up to 10 km/s. 
 
\subsection{Statistical analysis}

In this section, we present correlations between Mg~\textsc{ii} k physical parameters, as well as between those and the Si~\textsc{iv} derived from Paper I and refined as explained in paragraph~\ref{si iv fitting}. We used the RB asymmetry (introduced in \ref{velocities}), the peak separation, and the core depth.
As mentioned in \ref{fitting of mg profiles}, we excluded Mg \textsc{ii} k profiles with RL R$^{2} < 0.85$, as well as profiles with negative core depth (0.4$\%$), resulting from fitting failures.

The peak separation \citep{Pereira_2013} is a measure of the difference between the peak wavelengths, defined as 
\(\delta\lambda = \lambda_{2r} - \lambda_{2v}\). 
The core depth \citep{Schmit_2015} describes the depth of the k$_3$ intensity I$_3$ relative to the k$_2$ peak intensities I$_{2v}$ and I$_{2r}$: \(
D_k = 1 - 2I_3/(I_{2v}+I_{2r})\). A core depth equal to 1 means a profile with the deepest k$_3$ absorption, while a core depth equal to 0 a profile that is not self-reversed. The above parameters were calculated for the loop and jet masks, as well as a background region for comparison, as defined in paragraph~\ref{velocities}. We excluded the bright redshifted structure, as the Mg~\textsc{ii} k profiles there lack a clear central reversal, and thus the calculation of the above parameters may be unreliable. 
 
To quantify the relationships, we used the Spearman rank correlation coefficient, which measures the strength of monotonic associations between variables \citep{Spearman_1904}. This method is preferred over Pearson correlation in our analysis because the relationships between the parameters may be nonlinear or influenced by small sample sizes, and Spearman’s is more robust to these conditions.

\subsubsection{Correlations of Mg~\textsc{ii} k parameters} \label{mg correlations}
 
\begin{figure}[hbt!]
    \centering
     \resizebox{\hsize}{!}{\includegraphics{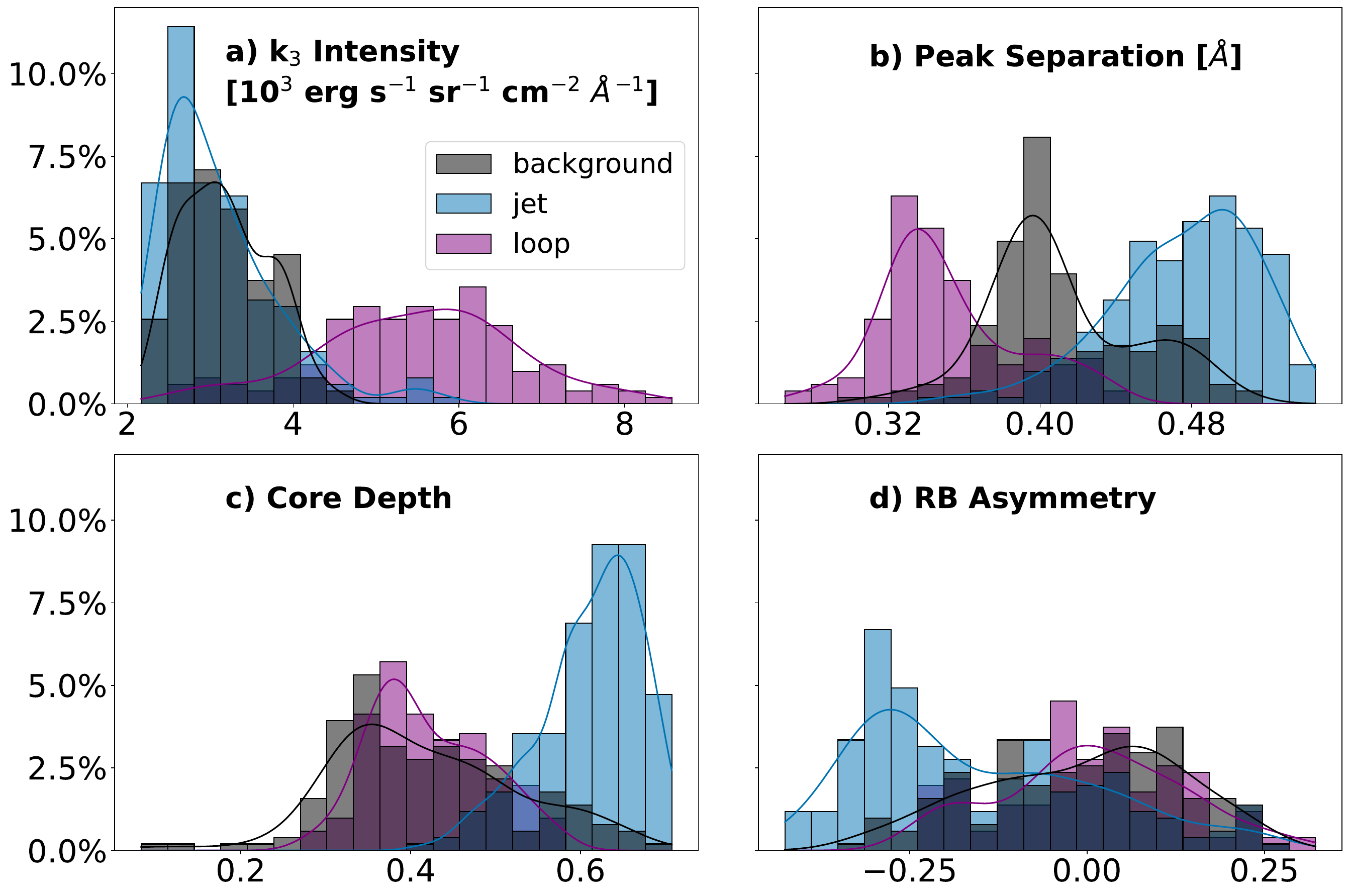}}
    \caption{Histograms of Mg~\textsc{ii} k parameters with a KDE overlay.
    }
    \label{mg2_distrib}
\end{figure}

\begin{figure}[hbt!]
    \centering
    \resizebox{\hsize}{!}{\includegraphics{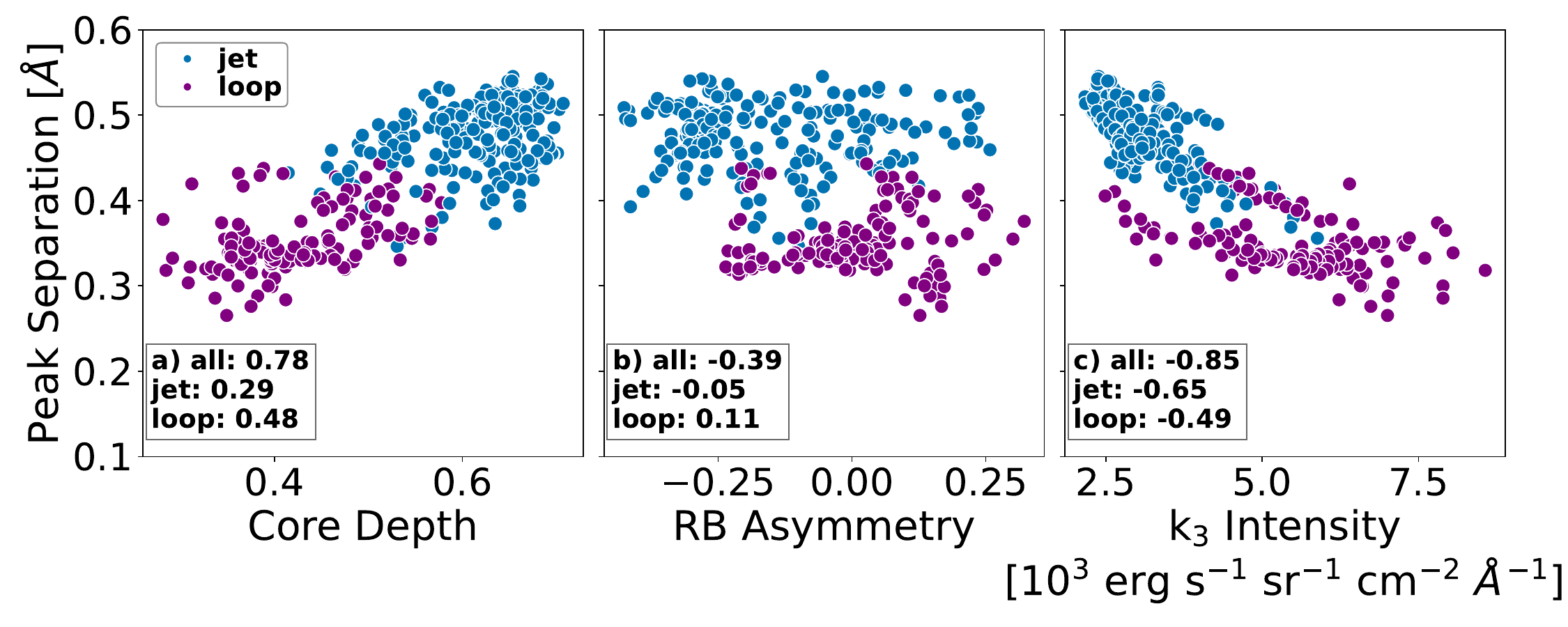}}
    \caption{Scatter plots of  Mg~\textsc{ii} k parameters,
    with the Spearman correlation coefficients for all scatter points and each structure shown individually. 
    }
    \label{fig:mg2_scatter}
\end{figure}

\autoref{mg2_distrib} presents the histograms of Mg \textsc{ii} k parameters for each region: k$_3$ intensity, peak separation, core depth and RB asymmetry. The  kernel density estimate (KDE) \citep{Rosenblatt_1956_kde} is over-plotted, in order to give a smooth estimate of the underlying distribution. The bandwidth is determined by Scott’s rule\footnote{\url{https://seaborn.pydata.org/generated/seaborn.kdeplot.html}},\footnote{\url{https://docs.scipy.org/doc/scipy/reference/generated/scipy.stats.gaussian_kde.html}} \citep{Scott1992MultivariateDE}.
Additionally, \autoref{fig:mg2_scatter} shows the scatter plots of the peak separation versus the core depth, RB asymmetry and k$_3$ intensity. As previously mentioned, negative and positive RB asymmetry values correspond to blueshifts and redshifts, respectively.

To assess differences in the distributions of the Mg~\textsc{ii} k parameters between regions, we applied the non-parametric Kruskal--Wallis H test \citep{Kruskal_1952}, which evaluates whether the samples from the different regions are drawn from the same distribution without assuming that the data are normally distributed.
Following, a Dunn's post-hoc test \citep{Dunn_1964} is applied for pairwise comparisons when significant ($p<0.05$). The results are summarized in \autoref{stat_tests} and confirm statistically significant differences (denoted with $\surd$ in \autoref{stat_tests}) across all measured parameters between the jet and the background, with the exception of the k$_3$ intensity. 

For the region below the loop, the core depth and RB asymmetry show no significant differences from the background. However, this region exhibits higher k$_3$ intensities and lower peak separations (see \autoref{mg2_distrib}). The jet is blueshifted, as indicated by the RB asymmetry, with
higher peak separations and core depths. 
We note that the mini-filament-like structure region, estimated roughly by its low k$_3$ intensity (green arrow in \autoref{rb maps}) and the blueshifted stripe in \autoref{inversion_maps}), is not statistically different from the background in any parameter, except for the k$_3$ intensity itself.

\begin{table}[t]
\caption{Statistical comparison of spectral parameters.}
\label{stat_tests}
\centering
\begin{tabular}{
  p{0.16\textwidth}
  p{0.08\textwidth}
  p{0.08\textwidth}
  p{0.08\textwidth}
}
\hline\hline
Feature & BG-jet &
BG-loop & Jet-loop \\
\hline
k$_3$ intensity & $\times$ & $\surd$ & $\surd$ \\
Peak separation & $\surd$ & $\surd$ & $\surd$ \\
Core depth & $\surd$ & $\times$ & $\surd$ \\
RB asymmetry & $\surd$ & $\times$ & $\surd$ \\
\hline
\end{tabular}
\tablefoot{Results are based on pairwise Dunn's post-hoc tests. A checkmark ($\surd$) indicates a significant difference ($p~<~0.05$), while a cross ($\times$) indicates no significant difference. BG refers to the background.}
\end{table}

The peak separation is correlated with the core depth and the k$_3$ intensity
for both structures (\autoref{fig:mg2_scatter}). Correlations within individual structures are weaker, suggesting that the overall stronger correlations mainly reflect differences between the two structures.
The correlations between the peak separation and RB asymmetry are almost zero for each structure separately. The slightly larger overall correlation highlights the distinction between the jet and the region below the loop and background, with the latter two being indistinguishable in this parameter. These relationships are analyzed in detail in \autoref{discussion}. 
We emphasize that these results should be interpreted with caution, as statistics do not necessarily imply physical relationships. However, they may reflect similar behavior at the atmospheric heights probed by these diagnostics.

\subsubsection{Correlations between Mg~\textsc{ii} k and Si~\textsc{iv} 1393~\AA\ parameters}
\label{correlations si mg}

\begin{figure}[hbt!]
    \centering \resizebox{\hsize}{!}{
\includegraphics{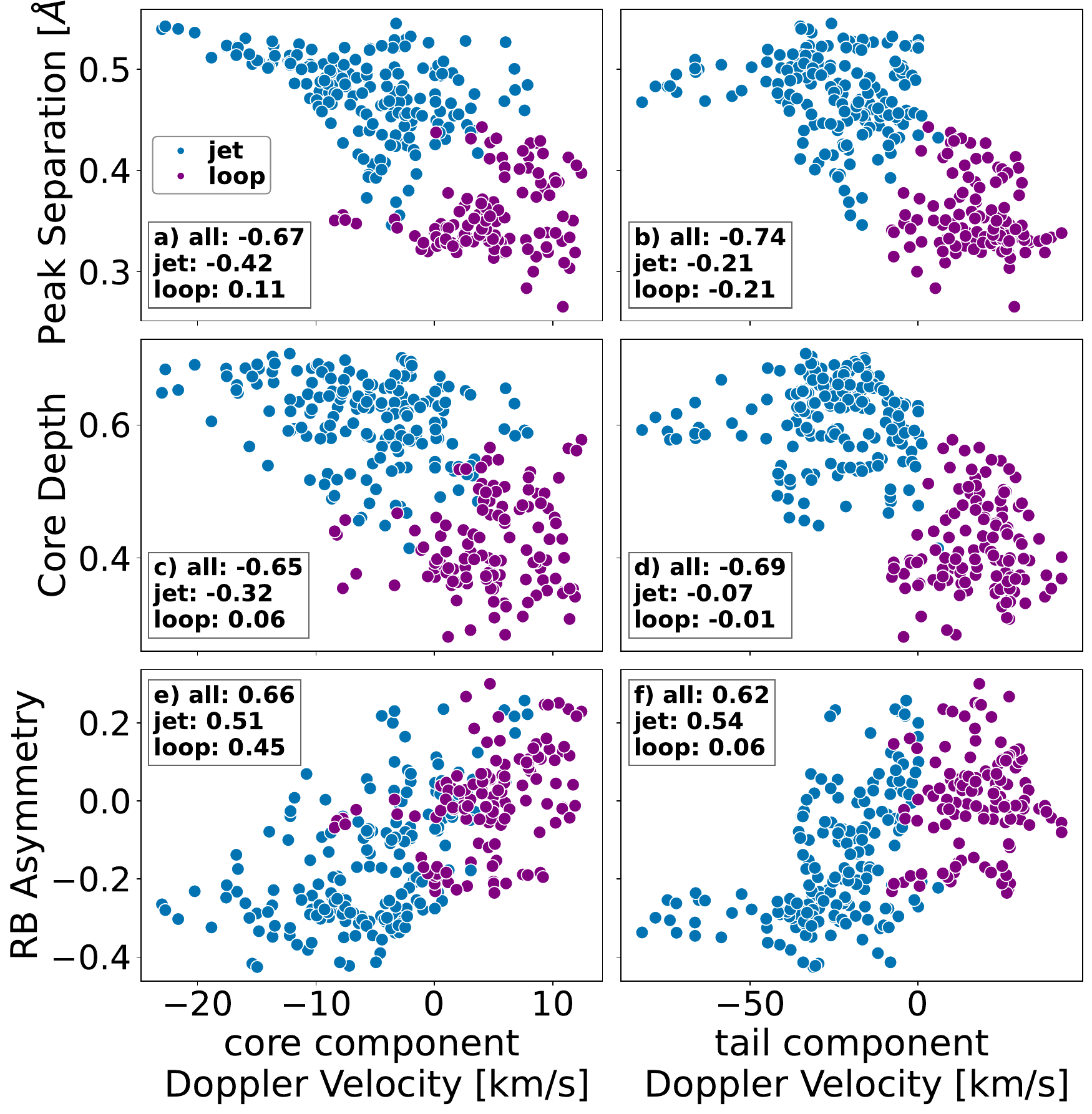}}
    \caption{
    Scatter plots of the Si~\textsc{iv} 1393~\AA\ Doppler velocities of the core and tail components versus Mg~\textsc{ii} k parameters. The Spearman correlation coefficient is also shown at each panel for all points and each structure separately.
    }
    \label{mg_si_scatter}
\end{figure}

\autoref{mg_si_scatter} shows the scatter plots between the Mg~\textsc{ii} k parameters- peak separation, core depth, and RB asymmetry- and the Doppler velocities of the Si~\textsc{iv} 1393~\AA\ line for the core and tail components, which are derived from the bimodal Gaussian fitting performed to the Si~\textsc{iv} profiles. The tail component is characterized as wider, with generally lower intensities and higher Doppler shifts, while the core component is narrower with higher intensities (see Paper I).

In all parameters, the jet generally presents higher correlations than the region below the loop.
The correlations are higher when taking into account both structures together, an effect that mostly highlights the clear distinction between them in these parameters, rather than point-wise correlations. 
The jet is characterized by Mg~\textsc{ii} k profiles with deep central reversals, large peak separations and blueshifted 
Si~\textsc{iv} velocities, indicating the connection between the chromospheric and transition region signatures of the cool jet component. On the other hand, the loop exhibits redshifted Si~\textsc{iv} velocities, associated with increased k$_3$ intensities and smaller peak separations of the Mg \textsc{ii} k profiles probing the region below it. 

More specifically, the positive correlations between the core and tail Doppler velocities and the RB asymmetry in the case of the jet indicate similar plasma motions between the upper chromosphere and transition region. The  zero correlation between the RB asymmetry and the Si~\textsc{iv} 1393~\AA\ tail component Doppler shifts 
within the loop region (panel~f) suggests that the event causing the  tail component profiles does not have an impact on the chromosphere, while the core component shows a stronger correlation. However this relationship may reflect a general behavior, since the loop is not distinguishable from the background in the RB asymmetry.

Generally, the three Mg~\textsc{ii} k parameters were calculated by combining intensities originating from the middle chromosphere with the core intensity forming in the upper chromosphere. 
Because the middle chromosphere is only weakly coupled to the transition region, parameters that mix contributions 
from both layers may naturally exhibit weaker correlations with transition region diagnostics, typically below 
50\%. Moreover, the loop is not directly visible in upper chromospheric temperatures, as it is formed in a temperature range from the low transition region to the corona, in contrast to the jet, which includes a cool counterpart. Thus, the correlations for the loop are expected to be weaker or nonexistent.

\section{Discussion} \label{discussion}

\begin{table*}[hbt!]
\centering
\caption{Characteristic events describing the mini-filament-like structure and jet evolution.}
\begin{tabular}{p{0.25\linewidth} p{0.72\linewidth}}
\hline\hline
Time (UT) & Event \\
\hline 
15:58 -- 16:58 & Converging motions and decrease of magnetic flux at PIL1  \\
$>$17:17 & Appearance of the mini-filament-like structure \\
17:20:53 -- 17:23:31 & Downward flowing plasma detected from the southern end of the mini-filament-like structure \\
17:23:43 & Upward motions and length increase of the mini-filament-like structure \\
17:24:07 & Appearance of elongated bright structure in AIA 304~\AA \\
17:24:55 & Brightening in AIA 304~\AA \\
17:25:19 & Appearance of the cool jet counterpart in AIA 304~\AA \\
17:26:30 & Hot jet counterpart and loop first appearance in higher-temperature AIA channels \\
\hline
\end{tabular}
\label{tab:events_table}
\end{table*}

In this section, we discuss the observed spectral signatures, the evolution of the mini-filament-like structure, and the magnetic topology of the region. We interpret our findings in the context of other studies in the literature and we examine how these observations can provide insights into the formation and evolution of the jet. 

\subsection{Spectral signature}

First of all, we observe a positive correlation between the peak
separation and core depth (see paragraph~\ref{mg correlations}), which was also found by \cite{Schmit_2015}. The enhancement of the peak separation can be explained by the effect of opacity broadening \citep{Rathore_2015a}, where
an increased opacity, corresponding to a higher column mass above the $k_2$ peaks formation height, results in a wider absorption profile, shifting the $\tau_\nu$ = 1 of the k$_2$ peaks to larger frequency offsets \citep{Hansteen_2023}. The higher opacity may also raise $\tau$(k$_3$) to larger altitudes, where the source function is lower, so that the k$_3$ intensity is decreased, leading also to an increased core depth. 

Especially on the jet spire, we find the largest peak separations and core depths, and the lowest k$_3$ intensities (Figures~\ref{mg2_distrib} and~\ref{fig:mg2_scatter}). There, the upward plasma motions, measured in the 
RB asymmetry (\autoref{rb maps}), may increase the column mass by transporting dense chromospheric plasma to larger heights, thereby enhancing the opacity. The variation in k$_2$ formation height is possibly implied by the observed redshifted velocities obtained by IRIS$^2$ at the jet base for $\log\tau\,=\,-4.8$ (\autoref{inversion_maps}), as well as $-4.8<\log\tau \leq\ -4$ (not shown). Furthermore, the deeper central reversals, as well as the lower $k_3$ intensities on the jet spire (\autoref{mg2_distrib}), suggest an elevated  $\tau(k_3)\,=\,1$ height found in numerical simulations  (see \cite{Leenaarts_2013b, Ondratschek_2024,ondratschek_2026} and \cite{Rathore_2015b} for the C \textsc{ii} lines). While the relationships reported in the above studies are weak and show substantial scatter, they may still indicate an underlying trend.

Furthermore, the correlations of the Si~\textsc{iv} velocities with both the RB asymmetry and the peak separation suggest that velocity gradients may provide an additional mechanism responsible for the enhanced peak separation. Simulations by \cite{Leenaarts_2013b,ondratschek_2026} indicate that the peak separation is wider in the presence of larger velocity gradients, as the latter cause wavelength shifts in different heights. According to the same authors, an exception is the case of a wider absorption profile in the presence of a temperature maximum in the lower chromosphere.
Another parameter that is believed to play a role in increasing the width of the absorption profile, thus enhancing the peak separation, is microturbulence \citep{Carlsson_2015,Hansteen_2023}. However, its exact nature and contribution to the formation of chromospheric spectral lines is still debated \citep{Judge_2020}. Besides, in most cases models cannot fully reproduce the Mg \textsc{ii} h$\&$k profiles and microturbulence is added as an ad hoc parameter (see \cite{Pereira_2013} and \cite{Bjorgen_2018} for the Ca \textsc{ii} H\&K lines). 

Generally, the clear distinction between the region below the loop and the jet in all the above parameters demonstrates that the jet and loop leave distinct chromospheric and transition region signatures. The jet exhibits strong dynamical effects detectable in both Mg~\textsc{ii} and Si~\textsc{iv}, while the loop doesn't influence the chromosphere in such extent, confirming that loops in general are coronal/transition region structures observed in higher temperatures.

\subsection{Mini-filament-like structure evolution}

\autoref{tab:events_table} shows the timeline of events leading to the jet eruption. We observe a mini-filament-like structure (arrows A and B in \autoref{all images}), which gets elongated and destabilized, possibly triggering the jet eruption. The mini-filament-like structure is initially observed very close to the jet base, similarly to other observations \citep{Sterling_2015}.

Before the appearance of the mini-filament-like structure, we observe possible weak, partial flux cancellation at PIL1
(see~\autoref{fig:hmi 720} and online HMI animation). 
Several recurrent increases in AIA 304~\AA\ intensity are also observed above PIL1, presumably resulting from reconnection driven by cancellation. This phenomenon has been reported in similar settings early in the cancellation phase \citep{Bura_2025} or prior to the mini-filament formation \citep{Panesar_2017}. \cite{Kontogiannis_2020} observed such increases in intensity before the mini-filament eruption, however without evidence of flux cancellation.

We also observe flux emergence and shear motions, which, along with flux cancellation, have been associated with the formation of magnetic arcades capable of supporting filament plasma, as they can increase the non-potentiality of the overlying magnetic configuration \citep{Ballegooijen_1989,Zhou_2025}. Moreover, flux cancellation has also been associated with mini-filament eruptions, as it can facilitate the internal reconnection below it and lead to its destabilization \citep{Panesar_2018}. 

Upward motions and elongation of mini-filaments, similar to our mini-filament-like structure, have been reported by other studies, for example \cite{Panesar_2017} but in larger time-scales.   
The estimated lifetime of 5 to 8 minutes, given the uncertainty in the exact formation time, and the maximum length of 11 Mm, are both within the ranges reported for mini-filaments in previous studies \citep{Sterling_2015,Chen_2025_mf}.
Moreover, the mini-filament-like structure is partially sampled by the IRIS slit at its rising motion, seen as a faint structure with low k$_3$ intensities (see \autoref{rb maps}) as well as blueshifted velocities of 4 km/s, low densities and temperatures derived from IRIS$^2$ at $\log \tau \,=\, -4.8$ (\autoref{inversion_maps}).

The evolution of the mini-filament-like structure is connected with two distinct structures in different timestamps. 
First of all, the IRIS slit captures a bright redshifted structure, spatially coinciding with the mini-filament-like structure prior to its rising (\autoref{all images}, \autoref{inversion_maps} and \autoref{rb maps}). The high intensity seen in the Mg~\textsc{ii} k spectra and 2796~\AA\ SJI implies that it exhibits higher temperatures than its surroundings, in accordance with the higher temperature derived from IRIS$^2$.
The inversions also show increased electron densities with respect to the surrounding plasma, as well as redshifted velocities. Although the latter originate from a not well-constrained optical depth of $\log\tau\,=\,-5.6$, they are consistent with the redshift derived qualitatively from the RB asymmetry. 

This structure may  represent drainage of mini-filament material along magnetic field lines from its southern footpoint rooted in a region of negative magnetic polarity, associated with destabilization of the mini-filament. 
\cite{Kontogiannis_2020} report similar results, as downflows are detected at one footpoint of a mini-filament. 
Plasma drainage has also been observed and studied in large-scale filaments \citep{cabezas_2024,Wu_2025}, where plasma heating at the site of drainage is caused by transformation of kinetic to thermal energy \citep{Gilbert_2013}. 

Another possibility is 
that the bright redshifted structure is a result of magnetic reconnection between falling mini-filament plasma and pre-existing loops, assuming that the features marked by arrow D in \autoref{all images} correspond indeed to low-lying loops. In addition, the high temperature further supports the possibility of localized heating due to reconnection \citep{Gilbert_2013}. Even if the mini-filament-like structure is not actually a mini-filament, the bright redshifted structure can still indicate plasma flowing downward following the structure's eruptive motion.

Nonetheless, all above scenarios are consistent with the fact that almost or fully single-peaked profiles dominate this region (\autoref{random_cool_profiles}). Various studies show that single-peaked profiles can be produced in flares and plages due to increased densities and temperatures  \citep{Carlsson_2015, Kerr_2015, Panos_2018}, as well as due to increased velocities at the line center \citep{Costa_Kleint_2017}. While the magnitude of these conditions differs from that in our observations, we do find increased temperatures and densities. Velocity gradients can also alter the shape of the profiles to such extent that the red peak is almost extinct, leading to an absent central reversal. To our knowledge, our work is one of the few studies observing such profiles associated with mini-filaments and jets.

 The second structure associated with the evolution of the mini-filament-like structure is an elongated feature, visible in AIA 304~\AA. It forms parallel to the mini-filament, develops concurrently with its rise (\autoref{all images}, arrow E1) and probably consists of multiple smaller-scale structures (\autoref{all images}, arrow E2).
A similar elongated bright structure was reported by \cite{Madjarska_2022}, who interpreted it as newly reconnected loops developing above the PIL and associated with quasi-separatrix layer (QSL) reconnection during the evolution of a micro-flare. In our case, it may represent loops associated with the destabilization and eruption of the mini-filament, forming after internal reconnection below it \citep{Li_2023}, in the case that a mini-filament is indeed present. However, we emphasize that this interpretation remains speculative.

\subsection{Magnetic topology and reconnection}

Generally, the potential field extrapolation provides evidence that magnetic reconnection may play a role in the mini-filament-like structure evolution and jet triggering (\autoref{fig:hmi 720_extrapolation}), as we find a null point at a height of 2 Mm in close proximity to the mini-filament's northern edge and a QSL close to the base of the jet between opposite sign polarities, serving as a favorable location for reconnection \citep{Li_2025,Guo_2013}. Null point reconnection has been associated with the eruption of mini-filaments and jets \citep{Cheng_2023,Kumar_2019}, as well as large-scale filaments \citep{Devi_2025,Li_2017}.
Additional indications of reconnection at the jet base include redshifted velocities derived from the IRIS$^2$ inversion (\autoref{inversion_maps}-arrow A), as well as the appearance of a brightening in AIA 304~\AA\ at the same location before the jet eruption (see online animation).

The jet consists of two components, with a cooler counterpart preceding the hotter one by approximately 1 to 2 minutes (arrows F1 and F2 in \autoref{all images}, respectively). Hot jet counterparts are often observed before the cooler ones \citep{Jiang_2007}, with the latter associated with mini-filament eruptions \citep{Moore_2010}. However, \cite{Shen_2017} report the opposite sequence, in which the cooler component appeared before the hotter counterpart. They suggest either that reconnection occurred at lower chromospheric heights or that the hot plasma resulted from heating of the cooler material. Both interpretations are broadly consistent with our observations, since we detect signatures of the jet in chromospheric diagnostics and a magnetic null point at the upper chromosphere. 

A possible mechanism for the formation of the jet and the associated loop is interchange reconnection between the rising mini-filament-like structure and the overlying open magnetic field lines \citep{Sterling_2015,Panesar_2016}. 
The potential field extrapolation topology reveals closed magnetic field lines spatially coincident with the loop and open magnetic field lines above the rising mini-filament-like structure, thereby providing a configuration favorable for this scenario.

\section{Conclusion} \label{conclusion}

In this work, we presented a comprehensive analysis of a solar jet,
within a low-latitude CH, combining diagnostics from the Mg~\textsc{ii} k and Si~\textsc{iv} spectral lines. By merging IRIS and SDO observations, we achieved a multi-layer study of the jet. 
The spectral analysis of Mg~\textsc{ii} k and Si~\textsc{iv} spectra diagnostics revealed a coupling between the chromospheric and transition region dynamics and plasma motions. Although the jet is not directly visible in the Mg~\textsc{ii} k intensity and the SJ 2796~\AA\ images, there is a signature of the jet in the shape and behavior of the Mg \textsc{ii} k profiles. The increased opacity caused by the jet plasma results in increased peak separations and core depths of the Mg \textsc{ii} k profiles. 

The jet eruption is also associated with a small-scale absorption structure, most likely a mini-filament, whose formation is preceded by magnetic flux emergence, shear motions, and weak flux cancellation in the photosphere. The mini-filament-like structure follows upward motions, accompanied by downward flows at its footpoint, possibly due to drainage. There, we find almost or fully single-peaked Mg~\textsc{ii} k profiles as a result of enhanced thermodynamic conditions or plasma motions. 

Although the exact mechanism responsible for the jet eruption cannot be fully elucidated, reconnection likely plays a role in triggering the evolution of the system, evident by the presence of a null point roughly at the upper chromosphere and a QSL below the jet base. The appearance of the cooler jet component earlier than the hotter one, along with the location of the null point and the chromospheric response of all structures associated with the jet eruption, leads us to the conclusion that the origin of the jet can be traced down to the chromosphere.

Overall, this study provides insight into the genesis of the jet and the chromosphere -- transition region connection.  
A novelty of this work is the use of GA and RL for the fitting of the Mg~\textsc{ii} and Si~\textsc{iv} profiles, making the fits more robust against noise and spectral profile variability. 
The expansion of this work to more observations is key in determining the generality of the above conclusions.
Unfortunately, the size of the structures under study is close to the resolution limit of the related instruments, making the analysis of this kind challenging. Nevertheless, such detailed multiwavelength observations of this kind can pose a significant challenge for MHD models of jets.

\begin{acknowledgements} 
We thank the anonymous referee for insightful comments and suggestions that improved this work. We would also like to thank J.-C. Vial for useful discussions. C.G. acknowledges
support 
by the Sectoral Development Program (O$\Pi\Sigma$ 5223471) of the Ministry of Education, Religious Affairs and Sports, through the National Development Program (NDP) 2021-25. S.P. acknowledges
support by the ERC Synergy Grant ‘Whole Sun’ (GAN: 810218). 
\end{acknowledgements}
\bibliographystyle{aa.bst} % style aa.bst
\bibliography{references.bib}

\end{document}